\documentclass[12pt]{article}
\usepackage{amssymb,amsmath,amsfonts,eurosym,geometry,ulem,graphicx,caption,color,setspace,comment,footmisc,caption,natbib,pdflscape,subfigure,array,tikz,hyperref, xcolor}

\usepackage[T1]{fontenc}
\usepackage[thinlines]{easytable}
\usepackage{enumerate}
\usepackage{amsthm}     
\usepackage{bbm}
\usepackage{bm}         
\usepackage{accents}
\usepackage{dcolumn}
\usepackage{color, colortbl}
\usepackage{tikz}
\usepackage{pgfplots}
\usetikzlibrary{shapes,decorations.pathreplacing, arrows.meta}
\usepackage{pgfplotstable}
\usetikzlibrary{arrows,automata}
\pgfplotsset{width=7cm,compat=1.8}
\usetikzlibrary{positioning,decorations.pathreplacing}
\usetikzlibrary{arrows,automata,patterns}
\usepackage{multirow}
\usepackage{dcolumn}
\usetikzlibrary{shapes.geometric}
\usepackage{multirow}
\usepackage{varioref}
\usepackage{array}
\usepackage{ulem}
\usepackage{booktabs}
\usepackage{threeparttable}

\makeatletter
\def\@footnotecolor{red}
\define@key{Hyp}{footnotecolor}{%
 \HyColor@HyperrefColor{#1}\@footnotecolor%
}
\def\@footnotemark{%
    \leavevmode
    \ifhmode\edef\@x@sf{\the\spacefactor}\nobreak\fi
    \stepcounter{Hfootnote}%
    \global\let\Hy@saved@currentHref\@currentHref
    \hyper@makecurrent{Hfootnote}%
    \global\let\Hy@footnote@currentHref\@currentHref
    \global\let\@currentHref\Hy@saved@currentHref
    \hyper@linkstart{footnote}{\Hy@footnote@currentHref}%
    \@makefnmark
    \hyper@linkend
    \ifhmode\spacefactor\@x@sf\fi
    \relax
  }%
\makeatother

\hypersetup{footnotecolor=black}

\hypersetup{
    colorlinks,
    linkcolor={black},
    citecolor={blue!50!black},
    urlcolor={blue!80!black}
}

\newcolumntype{L}[1]{>{\raggedright\let\newline\\\arraybackslash\hspace{0pt}}m{#1}}
\newcolumntype{C}[1]{>{\centering\let\newline\\\arraybackslash\hspace{0pt}}m{#1}}
\newcolumntype{R}[1]{>{\raggedleft\let\newline\\\arraybackslash\hspace{0pt}}m{#1}}
\newcolumntype{H}{>{\setbox0=\hbox\bgroup}c<{\egroup}@{}}
\usepackage{setspace}

\newtheorem{assumption}{Assumption}{\bf}
{\bf}

\newcommand{\E}{\mathrm{E}}
\newcommand{\Cov}{\mathrm{Cov}}
\renewcommand{\P}{\mathrm{P}}
\renewcommand{\iota}{\vec{\mathrm{1}}}
    \def\independenT#1#2{\mathrel{\setbox0\hbox{$#1#2$}%
    \copy0\kern-\wd0\mkern4mu\box0}} 
    
\usepackage{authblk}

\title{Testing the effects of an unobservable factor: \\ Do marriage prospects affect college major choice?\footnote{We are grateful for the valuable feedback from seminar participants at the National University of Singapore, the University of Notre Dame, the University of Pennsylvania, Mississippi State University, and Keio University. We also appreciate the input received during the 2023 Taiwan Economics Research Conference, the Conference in Celebration of Hashem Pesaran’s Achievements, and Econometric Society World Congress (Seoul, 2025).}}

\author[1]{H. Alper Arslan}
\author[2]{Brantly Callaway}
\author[3]{Tong Li}

\affil[1]{Department of Economics, University of Texas at San Antonio}
\affil[2]{Department of Economics, University of Georgia}
\affil[3]{Department of Economics, Vanderbilt University}

\date{\today}

\begin{document}

\begingroup
\setstretch{1.4}                 
\maketitle
\thispagestyle{empty}            
\vspace{-2.5em}                  

\begin{abstract}
\noindent This paper develops an econometric test for whether an unobservable factor jointly influences a choice among multiple alternatives and a subsequent binary outcome. Because only transformed versions of the choice model's unobservables are identified, we combine a copula-based joint model with results on the covariance between ordered and unordered normal variables to recover the covariance between structural shocks, yielding a simulation-based test for a common unobserved factor. Applying it to the National Longitudinal Survey of Youth 1997, we find unobserved marriage prospects significantly influence college major choice. Supporting this, the dependence disappears once stated marriage expectations, elicited before major selection, are included.
\end{abstract}


\endgroup

\setcounter{page}{0} \thispagestyle{empty}\newpage

\section{Introduction}
\normalsize

Preferences, beliefs, attitudes, expectations, and valuations are critical elements of many economic models and play significant roles in agents' decisions. However, these types of variables are usually not observable in the available data. The econometrics literature has developed various methods (e.g., instrumental variables, control function approaches) to estimate the effects of observable factors on an outcome variable in the presence of unobserved factors. While these estimation methods yield useful insights in some contexts, they cannot be universally applied to understand the impacts of unobservable factors.

This paper focuses on analyzing the effects of an unobservable factor  (i.e., a confounding variable)  in a scenario involving a polychotomous choice (unordered multinomial choice) and a binary dependent outcome variable. To illustrate the idea, consider the following example related to students' choices of college majors. In the National Longitudinal Survey of Youth (97) dataset (NLSY97), it is evident that college graduates with different majors have considerable differences in marital status during their 30s. This observation raises two questions: (i) does the chosen major impact the marital outcomes of college graduates? and (ii) do students make their major decisions based on the marriage prospects (preferences, expectations) associated with different college majors? While the econometrics literature provides several methods to study the first question — which can be viewed as a treatment effect problem (e.g., \citet{heckman2006earnings}, among many others)— there is no straightforward method that satisfactorily answers the second. In this context, the unobservable ``marriage prospects'' factor may influence both marriage outcomes and college major choices. Researchers typically have access to data on college major choices, marital status, and a rich set of control variables related to the socioeconomic background. Yet, it remains unclear whether such data can reveal whether unobserved marriage prospects indeed influence students’ decisions regarding their choice of college major.

To address these questions, we develop an econometric framework that allows unobserved factors to jointly influence both a polychotomous choice and a subsequent binary outcome. Our main methodological contribution is a formal test for the presence of such common unobservable factors. The framework extends the binary endogenous choice models of \citet{evans1995finishing} and \citet{altonji2005evaluation} to settings in which the endogenous explanatory variable is polychotomous. It also builds on the insights of \citet{lee1983generalized} and \citet{dahl2002mobility}, who generalize the Heckman selection model from binary to polychotomous selection environments. Within this framework, we model the joint dependence between the unobserved components of the multinomial choice model and the binary outcome equation using a copula-based specification, following the methodology of \citet{chen2006estimation}, \citet{arellano2017quantile}, and \citet{callaway2017quantile}. This specification allows us to separate the identification of the causal effects of the polychotomous choice on the outcome from the dependence structure induced by unobserved confounding factors. By exploiting the ordering structure of latent utilities in the polychotomous choice model, we construct a test for whether the unobserved factors influencing the binary outcome are systematically related to the latent utilities that determine the choice among alternatives.

While our econometric model leads to a direct identification of the effects of polychotomous choices on binary outcomes (the first question above), identifying the dependence structure between the unobservable components of the two equations (the second question above) confronts
an additional challenge. The difficulty arises from the standard normalization required for identification in any polychotomous choice model: in our framework, latent utilities are normalized by taking differences between the highest and second-highest utilities, so only
transformed versions of the unobservable components in the choice model are identified. The dependence of interest, however, lies between the outcome disturbance and the \emph{untransformed} latent utility shocks---in our empirical context, the unobservable marriage-prospect factor is embedded in these primitive shocks. As a result, the
dependence parameters estimated in our copula framework capture rank-based dependence between the transformed unobservables rather than the structural covariances of interest. To bridge this gap, we build on results of \citet{siegel1993surprising} and \citet{rinott1994covariance}
characterizing the covariance between ordered and unordered components of a multivariate normal vector. These results imply that, under joint normality, the absence of structural dependence between the binary-outcome shock and the latent utility shocks implies zero
covariance between the outcome disturbance and the normalized latent utility difference arising in the choice model---a reduced-form implication of the structural null that is testable using the objects identified in our framework.

Because the normalized latent utility difference is generated through a selection operator and is therefore generally non-normal, the copula dependence parameter estimated in our model does not directly equal the structural covariance of interest. We therefore develop a simulation-based calibration procedure that maps the copula dependence parameter into the implied covariance between the structural shocks using the estimated marginal distributions of the error terms. This procedure allows us to construct the null distribution of the covariance statistic under the hypothesis that the unobserved components of the choice and outcome equations are independent. Comparing the empirical statistic with this simulated distribution yields a direct test for the presence of common unobservable factors—such as marriage prospects—that jointly influence both college major choice and marital outcomes.

To implement our test, we examine how marriage prospects influence college major choice using data from the NLSY97. The dataset contains detailed information on participants’ demographics, marital status, educational background, earnings, work hours, and survey responses reflecting their expectations. Our results reveal a significant association between the unobservable factors affecting marital status and college major choice. We interpret these unobservable factors as marriage prospects,\footnote{To support our interpretation of the unobservable factors as marriage prospects, we exploit our access to some unique survey questions in the NLSY97 that elicit respondents’ self-assessed probabilities of marriage. When we include these responses in our analysis, our approach no longer detects a relationship between the unobservable factors in marriage and major choice decisions. This finding provides empirical support for interpreting the unobservable factors in our main results as marriage prospects.} and, thus, our results indicate that students take into consideration their marriage prospects when they decide on their college major.

The proposed test for the effect of unobservable factors is related to several other papers that consider polychotomous choice models with selection based on unobservables. Motivated by a different application, which is to estimate hospital quality in a model with a binary dependent variable (mortality) and non-random selection, \citet{geweke2003bayesian} consider a binary choice model with a potentially endogenous polychotomous hospital choice variable as one of the explanatory variables. They propose a Bayesian method using a Markov chain Monte Carlo posterior simulator to estimate the model. In contrast, we propose a frequentist approach to deal with a similar model and to provide a direct test for the effect of unobservable factors.\footnote{\citet{debtrivedi} consider count data models with selectivity, in which they explicitly model the endogeneity of the polychotomous variable as driven by some unobserved factor loadings that are normally distributed and also appear in the equation for the count outcome.} Notably, our test does not impose restrictions on the correlations between the outcome and latent utilities within the choice model, which is commonly used in the literature as a result of the normalization of latent utilities.\footnote{It is worth noting that our paper is also related to the work on estimating treatment effects in models featuring selectivity (e.g., \citet{heckman2004using} and \citet{heckman2006understanding}). Our estimation method provides an approach for estimating treatment effects when unobservable factors have an expected effect on the selection into treatment and the outcome. Consequently, our approach contributes to the toolbox of treatment effects estimation techniques in the presence of confounding variables.}

In the labor economics literature, dynamic choice models are widely used to analyze sequential decision-making processes. For instance, \citet{keane_wolphin_97}, \citet{eckstein1999youths}, and \citet{belzil2002unobserved} estimate structural dynamic models of educational choices and find that expected earnings have a significant positive effect on college enrollment. Similarly, \citet{arcidiocono_04, arcidiocono_05} examine the role of expected earnings in major selection, employing structural dynamic models with innovative methodological advances. Our approach complements this line of research by offering a direct econometric test of whether unobserved factors influencing the initial decision (e.g., college major, a polychotomous choice) are systematically related to unobserved factors affecting a subsequent decision (e.g., marital status, a binary outcome).

Finally, our empirical application offers new insights into the determinants of college major choice by providing observational evidence on the relationship between major choice and marriage outcomes. While prior studies have extensively documented the role of non-pecuniary factors in educational decisions, they have largely focused on college enrollment or abstract preferences. For example, \citet{beffy_12} show that non-pecuniary considerations significantly influence major choice among French university students, and \citet{wiswall_zafar_15} use experimentally elicited beliefs to highlight the role of heterogeneous tastes in driving major decisions. Similarly, \citet{wiswall2021human} find that students consider not only future earnings but also factors such as a potential spouse’s income and fertility preferences when evaluating human capital investments, with family expectations playing a particularly important role for women. \citet{ersoy2022opening} further demonstrate that students adjust their preferences when exposed to non-earnings-related information through a staggered intervention.
While a separate literature explores the connection between marriage and college participation—mostly in theoretical terms (e.g., \citet{iyigun_walsh_07}; \citet{chiappori_iyigun_weiss}; \citet{gousse2017marriage}; \citet{zhang2017marriage})—empirical evidence remains limited. Notably, \citet{ge_11} and \citet{ge2022elite} examine how marriage prospects affect college entry decisions for women as well as career and family outcomes, and \citet{bursztyn2017acting} find that women may alter behaviors associated with professional success to avoid penalties in the marriage market.\footnote{Additionally, a growing literature examines assortative matching in the marriage market based on college attended and field of study. See, for example, \citet{artmann2021field} and \citet{kirkeboen2025college}, among others.}
In contrast to this existing work, our study uses observational data to directly examine whether marriage prospects influence college major choice, offering a new perspective on how personal and family formation considerations shape educational decisions.

The organization of this paper is as follows. Section 2 describes our econometric model for the polychotomous choice and the binary outcome variables, as well as the marginal and joint modeling of these variables. Section 3 provides details on estimating our model. Section 4 presents the testing procedure for common or correlated unobservable factors (e.g., marriage prospects) affecting both the polychotomous choice and the binary outcome. Section 5 presents the data, summary statistics, and preliminary estimation results. In Section 6, we apply the proposed test to examine the effects of marriage prospects on college major choice and marital outcomes using NLSY97 data. Section 7 presents results separately for females and males. Section 8 concludes the paper.

\section{The Econometric Approach} \label{sec:2}

Our econometric model encompasses two distinct components: a polychotomous choice and a binary outcome. Notably, the binary outcome (choice) occurs after agents have experienced the consequences of their polychotomous choices. Our primary econometric objective is to model unobserved factors within the polychotomous choice framework and their potential influence on the binary outcome variable. In our empirical application, individuals select their college majors, represented as  a polychotomous choice, while their marriage decisions correspond to binary outcomes. Our central empirical focus is to investigate the importance of an unobserved marriage-prospect factor in influencing college major choices.

\subsection{Marginal models of polychotomous choice and binary outcome}

\noindent \textbf{Polychotomous choice.} Assume that a decision maker $i$ selects an option among $J$ alternatives. Within our model, decision maker $i$ derives $V_{ij}$ utility from choosing $j$. We conceptualize this utility as a composite of attributes associated with the chosen alternative, the decision maker, and the expected features of each choice. 

Following standard discrete choice models, the decision maker chooses the option that maximizes the expected utility. As a consequence, decision maker $i$ chooses alternative $j_i^*$, which gives the highest expected utility:
\begin{equation}
j_i^*= \underset{j \in \{1, \hdots, J\}}{\arg \max} \ V_{ij} \label{major}
\end{equation}
where $V_{ij} = x_{1i}\beta _{j} + z_j \delta  +u_{ij}$. $z_j$ captures the observable attributes of each alternative that play a role in the decision maker's choice. $x_{1i}$ represents decision maker $i$'s characteristics at the time of the choice, $\beta_{j}$ is the corresponding parameter vector associated with the alternative $j$. The term $u_{ij}$ accounts for the latent aspect of utility, encompassing all unobservable factors, including unobserved expectations and preferences. We assume that the latent utility $u$ remains independent of the observed characteristics $z, x_1$, which is a standard assumption in the discrete choice literature.

\bigskip

\noindent \textbf{Binary outcome variable.} Next, we consider a model for the binary outcome in our framework. We model the binary outcome as a function of the observed attributes of the chosen alternative and unobservable factors; in particular,
\begin{equation}
m_{i} = \mathbf{1}\Big(x_i \mathbf{\gamma} + \sum_{j=1}^J j_i\tau_j + \epsilon_{i} >0 \Big)
\label{marriage}
\end{equation}
where $m_i$  is an outcome indicator equal to one if the positive outcome is realized and zero otherwise. In Equation \ref{marriage}, $x_i$ represents the observed attributes of decision maker $i$. These attributes encompass observable individual characteristics, including those in the polychotomous choice equation ($x_1$). We also include additional variables as part of $x_i$---these variables differ from the characteristics in the polychotomous choice equation. Some variables enter the polychotomous choice only as expected quantities, but their realized counterparts replace them once the consequences of the choice are observed. This inclusion allows us to control for post-choice realizations after the polychotomous choice has been made.\footnote{For example, earnings and working hours are only able to enter polychotomous as expected variables, but after graduation realized earnings and working hours replace them. This inclusion allows us to control for post-choice realizations after the polychotomous choice has been made.}

The parameter $\tau_j$ represents the causal impact of chosen alternative $j$ on an individual's likelihood of obtaining a positive outcome—a focal point in numerous econometric models. Researchers may obtain indirect evidence on the role of unobserved selection by comparing estimates of $\tau_j$ from models that ignore unobserved dependence between choice and outcome with estimates from models that allow for such dependence. Our econometric procedure differs by providing a direct test of whether unobserved factors affecting the choice decision are systematically related to the binary outcome.

$\epsilon_i$ encompasses the latent components of the binary outcome equation, including the unobservable factors by design. We assume that $\epsilon$ is independent of individual characteristics $x$. This implies that the unobserved factors influencing the binary outcome are distributed identically across individuals with different observable characteristics  However, we refrain from imposing any assumptions regarding the relationship between alternative, $j_i$, and the unobservable factors $\epsilon_i$. If the decision maker's choice depends on a specific unobservable factor, $j_i$ becomes an endogenous variable.  Our model allows for this possibility, and testing this hypothesis is a primary objective of our paper.\footnote{In our empirical application, the binary outcome corresponds to the individual's marriage decision, and the relevant unobservable factor is a marriage-prospect factor that may be correlated with gender-specific marriage preferences. While we address this specific concern by separately estimating male and female students in our application, for theoretical discussions, we primarily focus on exogenous covariates.}

\subsection{Joint model of binary outcome and polychotomous choice}
\label{sec:jointmodeling}

Given our objective of assessing whether an unobserved factor influences the polychotomous choice, we jointly consider the models in Equations \ref{major} and \ref{marriage}. Joint estimation of these equations enables us to analyze the covariance matrix of the error terms associated with the binary outcome and the polychotomous choice models. Since the influence of unobservable factors is captured within these error terms, the structure of their covariance matrix is central to our subsequent analysis.

Towards this end, we define a variable $I_i$ based on the polychotomous choice model in Equation \ref{major}. This variable takes values from $1$ to $J$, and $I_i=j^*$ if alternative $j^*$ is chosen. Therefore, $I_i=j^*$ if and only if

\[
\left( x_{1i}\beta_{j*} + z_{j^*}\delta + u_{ij^*} \right) \geq \underset{j\in\{\{1, \hdots, J\} \setminus j^*\}} \max{y_{ij}},
\]
where $y_{ij}= x_{1i}\beta_{j} + z_{j}\delta + u_{ij}$ is the latent utility of chosen major $j$ for student $i$. Then,
\begin{equation}
\left(x_{1i}\beta_{j*} + z_{j^*}\delta \right) \geq \xi_{ij^{*}}  \equiv  \underset{j\in\{\{1, \hdots, J\} \setminus j^*\}} \max{y_{ij}} - u_{ij^*}
\label{polychotomous}
\end{equation}

Equation \ref{polychotomous} shows that we can represent a polychotomous choice problem with a single inequality, rather than keeping track of $J$ latent utility equations. This simplification further underscores that the error terms of multinomial choice models can only be inferred up to a transformation \citep{lee1983generalized,dahl2002mobility}.

To characterize the joint distribution of the unobserved components $(\epsilon,\xi)$, we apply Sklar's Theorem \citep{sklar1959distribution}, which implies that any joint distribution can be decomposed into its marginal distributions and a copula capturing their dependence. In particular,
\[
F(\epsilon,\xi)=C\big(F_1(\epsilon),F_2(\xi)\big),
\]
where $F_1$ and $F_2$ denote the marginal distributions of $\epsilon$ and $\xi$, respectively, and $C(\cdot,\cdot)$ is a copula function. Following \citet{lee1983generalized}, we assume a Gaussian copula, which yields

{\small
\begin{equation}
C\big(F_1(\epsilon),F_2(\xi)\big)
=
B\!\left(
\Phi^{-1}\!\big(F_1(\epsilon)\big),
\Phi^{-1}\!\big(F_2(\xi)\big)
\,\middle|\, \boldsymbol{\rho}
\right),
\label{eq:copula_normal}
\end{equation}
}

where $B(\cdot,\cdot|\boldsymbol{\rho})$ denotes the bivariate normal distribution with correlation vector $\boldsymbol{\rho}$.\footnote{As established in \citet{chernozhukov2023distribution}, there is a local Gaussian representation of the joint distribution of $\epsilon$ and $\xi$, in the sense that the covariance vector $\boldsymbol{\rho}$ in Equation \ref{eq:copula_normal} depends on the values of $\epsilon$ and $\xi$. Effectively, Lee (1983) assumes the global Gaussian representation, which we follow in (\ref{eq:copula_normal}) by assuming $\boldsymbol{\rho}$ is a constant.} Under this specification, the transformed variables
\[
Z_\epsilon=\Phi^{-1}(F_1(\epsilon)), \qquad
Z_{\xi_j}=\Phi^{-1}(F_2(\xi_j))
\]
are jointly normal with zero means, unit variances, and correlations $\rho_j$. The parameters $\rho_j$ therefore summarize the dependence between the transformed latent components of the binary outcome and the polychotomous choice models while allowing their marginal distributions to remain flexible.

Next, we demonstrate how the proposed model relates to the observed joint behavior. Specifically, the probability of not receiving a positive outcome and selecting alternative $j^*$ is given by:

{ \small
\begin{align}
\P(m_i=0, I_i=j^* | x_i, z) &= \P\left(\epsilon_i\leq -x_i\gamma - \tau_{j^*}, \xi_{ij^*} \leq  x_{1i}\beta_{j^*} + z_{j^*}\delta \Big| x_i, z\right)\nonumber \\[5pt]
 &= B\left(\Phi^{-1}\big(F_1(-x_i\gamma - \tau_{j^*})\big), \Phi^{-1}\left[F_2\left( x_{1i}\beta_{j^*} + z_{j^*}\delta \right)\right] \Big| \boldsymbol{\rho} \right)
\label{prob_0j}
\end{align}} and, using the same type of arguments, the probability of receiving a positive outcome and choosing alternative  $j^*$ is given by

{ \small 
\begin{align}
 \P(m_i=1,  & I_i=j^* | x_i, z) = \P(I_i=j^* | x_i, z) -  \P(m_i=0, I_i=j^* | x_i, z) \nonumber \\[5pt]
    & \hspace{10pt} = F_2\left(x_{1i}\beta_{j^*} + z_{j^*}\delta \right) - B\left(\Phi^{-1}\big(F_1(-x_i\gamma - \tau_{j^*})\big), \Phi^{-1}\left[F_2\left(x_{1i}\beta_{j^*} + z_{j^*}\delta \right)\right] \Big| \boldsymbol{\rho} \right)
    \label{prob_1j}
\end{align}
}
$\P(m_i=0, I_i=j^* | x_i, z_i)$ and $\P(m_i=1, I_i=j^* | x_i, z_i)$ characterize all possible binary outcomes for all alternative choices.  Equations \ref{prob_0j} and \ref{prob_1j} suggest that we can use these probabilities to estimate all the parameters of the joint distribution of $F(\epsilon,\xi)$, given our model for binary outcome and alternatives.

\subsection{Identification}

The parameters of the marginal models are identified using standard arguments for discrete choice and binary response models. Variation in $(x_{1i},z_j)$ identifies the parameters $(\beta,\delta)$ of the polychotomous choice model, while variation in $x_i$ identifies the parameters $(\gamma,\tau_j)$ of the binary outcome equation conditional on the realized choice.

The key identification challenge concerns the dependence parameters linking the unobserved components of the two equations. In our framework, this dependence is summarized by the copula parameters $\rho_j$, which capture the correlation between the transformed shocks $\Phi^{-1}(F_1(\epsilon_i))$ and $\Phi^{-1}(F_2(\xi_{ij}))$.

Identification of $\rho_j$ relies on exclusion restrictions. In particular, the alternative-specific variables $z_j$ affect the utility index of the polychotomous choice model,
\[
V_{ij} = x_{1i}\beta_j + z_j\delta + u_{ij},
\]
but do not enter the binary outcome equation except through the realized choice $I_i=j$. As a result, variation in $z_j$ shifts the probability of selecting alternative $j$ without directly affecting the latent outcome equation. This generates observable variation in the joint distribution of $(m_i,I_i)$ that allows the copula dependence parameters to be point identified.

Intuitively, if shifts in $z_j$ alter the probability of choosing alternative $j$ and simultaneously change the conditional probability of the binary outcome, this comovement must arise from dependence between the unobserved components of the two equations.

Importantly, identification of the copula parameters does not require joint normality of the primitive shocks $(\epsilon_i,u_{i1},\ldots,u_{iJ})$. The Gaussian copula specification only imposes a parametric structure on the dependence between the marginal distributions $F_1(\epsilon)$ and $F_2(\xi_j)$. Distributional assumptions become relevant only when interpreting the estimated dependence parameters in terms of the structural covariances $\Cov(\epsilon_i,u_{ij})$, which is the focus of the testing procedure developed in Section \ref{testing}.

\section{Estimation}
\label{sec:estimation}

In this section, we describe our estimation procedure, building on the discussion in Section \ref{sec:2}.  We proceed in two steps.  In the first step, we estimate the polychotomous choice model.  In the second step, we simultaneously estimate the parameters of the binary choice model and the copula parameters, given the estimated parameters from the first step.  
    
\subsection*{Step 1: Polychotomous choice model}

We first need to specify $F_2\Big(x_{1i}  \beta_{j^*} + z_{j^*}\delta \Big)$, where $F_2(\cdot)$ characterizes the error components in the polychotomous choice model to estimate the Gaussian copula presented in Equation \ref{eq:copula_normal}. In this step, we present a multinomial probit specification for polychotomous choice and binary outcome as well as a testing procedure. Alternatively, multinomial logit model can be used and reduce computational complexity, and we provide additional results based on using a multinomial logit model in Appendix \ref{section:appendixEst}.

While the latent utilities are normalized with the second highest latent utility of the choice in our econometric approach, a researcher does not know this choice without additional data on the ranking of alternatives. Therefore, we assume that all majors, except the chosen one, have a chance of being the second choice of the student. If we assume that $u_{ij},\; j=1, \hdots, J$ are distributed with jointly normal, $F_2\left(x_{1i}  \beta_{j^*} + z_{j^*}\delta  | \theta \right) = \P\left(\xi_{ij^*} < x_{1i}  \beta_{j^*} + z_{j^*}\delta  \right)$, where $\theta$ includes parameters such as $\beta$, can be represented by using all major options as:

\begin{align*}
F_2\Big(x_{1i}  \beta_{j^*} + z_{j^*}\delta  \Big| \theta \Big) = & \P\left(\xi_{ij^*} < x_{1i}  \beta_{j^*} + z_{j^*}\delta  \right) \\
=& \P\left( x_{1i}  \beta_{j^*} + z_{j^*}\delta + u_{ij^*}\geq x_{1i}  \beta_k + z_k \delta + u_{ik}, \; \forall k\neq j^*\right) \\
=& \P\left(\tilde{V}_{ikj^*} + \tilde{u}_{ikj^*} \leq 0  \; \forall k\neq j^*\right) \\
=& \int \mathbf{1}\left(\tilde{V}_{ikj^*} + \tilde{u}_{ikj^*} \leq 0  \; \forall k\neq j^*\right) \phi(\tilde{u}_{ij^*}) d\tilde{u}_{ij^*} \\
=& \int (\tilde{u}_{ij^*}\in A_{ij}) \phi(\tilde{u}_{ij^*}) d\tilde{u}_{ij^*}
\end{align*}

\noindent where $\tilde{V}_{ikj^*}= x_{1i} (\beta_k - \beta_{j^*}) + (z_k -z_{j^*})\delta $ and $\tilde{u}_{ikj^*}= u_{ik}- u_{ij^*}$. $A_{ij^*} \equiv \{ \tilde{u}_{ij^*} \; s.t. \;  \tilde{V}_{ikj^*} + \tilde{u}_{ikj^*} \leq 0  \; \forall k\neq j^*\}$. Since this probability is a $(J-1)$ dimensional integral over error differences $A_{ij^*}$, the subsequent MLE procedure becomes computationally intensive if the number of majors is large. This issue can be addressed by employing a simulation-based approach to handle the integrals that would otherwise be challenging to compute. However, the utilization of the simulation technique may lead to a nonsmooth objective function due to the index function. As a result, we use the GHK simulator (\citet{Geweke1992}, \citet{hajivassiliou1996simulation}, \citet{Keane1994}), which is widely regarded as the most reliable method for simulating multivariate normal rectangle probabilities (see, e.g., \citet{train2009discrete}).\footnote{We provide additional details of the GHK simulator in Appendix \ref{section:appendixEst}.}

\subsection*{Step 2: Simultaneous estimation of binary choice and Gaussian copula}

Next, we derive and estimate $\P(m_i=0, I_i=j^* | x_i, z)$ and $\P(m_i=1, I_i=j^* | x_i, z)$ using  a copula function and $F_2\Big(x_{1i}  \beta_{j^*} + z_{j^*}\delta  \Big)$, which is available after the first step. Recall that we use the Gaussian copula for its practicality in characterizing the dependence of marginal distributions. Accordingly, using the equality in Equation \ref{prob_0j} and the Gaussian copula in Equation \ref{eq:copula_normal}, we can write $\P(m_i=1, I_i=j^* | x_i, z)$ as the following:\footnote{Details of this derivation are available in Appendix \ref{section:appendixEst}.} 

\begin{small}
\begin{align*}
\P(m_i=&1, I_i=j^* | x_i, z) =  \int_{- \infty}^{(-x_i\gamma- \tau_{j^*})} \Phi [ (\Phi^{-1}(F_2(V_{ij^*}))) + \rho_{j^*} \epsilon_i)/ (1-\rho_{j^*}^2)^{1/2}] \phi(\epsilon) d \epsilon 
\end{align*}
\end{small}

\noindent where $ V_{ij^*}=x_{1i}  \beta_{j^*} + z_{j^*}\delta $.\footnote{Note that we characterize $F_2\left(x_{1i}  \beta_{j^*} + z_{j^*}\delta  | \theta \right) = \P\left(\xi_{ij^*} < x_{1i}  \beta_{j^*} + z_{j^*}\delta  \right)$. Therefore, $F_2(V_{ij^*}) = \P\left(\xi_{ij^*} < x_{1i}  \beta_{j^*} + z_{j^*}\delta  \right)$}  Finally, we assume that the marginal distribution of $\epsilon_i$ follows a standard normal distribution consistent with a standard probit model. Then, $F_1(\cdot)=\Phi(\cdot)$, $\Phi^{-1}(F_1(\epsilon))= \epsilon$. Using these definitions, we can write the likelihood function as follows:

\begin{small}
\begin{align*}
 \log L(\theta) & =   \sum_{i=1}^N \sum_{j=1}^J \;  \{1(I_{i}={j^*})\; \log \overset{(-x_i\gamma - \tau_{j^*})}{ \underset{- \infty}{\int}} \Phi [ (\Phi^{-1}(F_2({V}_{ij^*})) + \rho_{j^*} \epsilon_i)/ (1-\rho_{j^*}^2)^{1/2}] \phi(\epsilon) d \epsilon \}^{(m_i=0)} \\
&  \{1(I_{i}={j^*}) \; \log [ F_2({V}_{ij^*}) - \overset{(-x_i\gamma -\tau_{j^*})}{ \underset{- \infty}{\int}}  \Phi [ (\Phi^{-1}(F_2({V}_{ij^*})) + \rho_{j^*} \epsilon_i)/ (1-\rho_{j^*}^2)^{1/2}]  \phi(\epsilon) d \epsilon ] \}^{(m_i=1)} 
\end{align*}
\end{small}

\noindent where $\theta=\{\beta, \delta, \gamma, \mathbf{\tau}, \rho \}$. Therefore, our econometric model can be estimated using maximum likelihood methods. The parameter estimates are obtained by solving the following optimization problem:
\[
\hat{\theta}=  \underset{\theta} \arg \max \  \sum_{i}^N \log L(\beta, \delta, \gamma,  \mathbf{\tau}, \rho)
\]
In practice, this estimation becomes a simulated maximum likelihood estimation (SMLE) if choice probabilities are simulated using the GHK simulator.

\section{Testing Structural Dependence via Copula Calibration}
\label{testing}

Our primary objective is to determine whether unobserved factors jointly influence the polychotomous choice and the subsequent binary outcome. In the framework developed above, this dependence is captured by the copula-based parameters
\begin{equation}
\rho_j 
=
\Cov\!\left(\Phi^{-1}(F_1(\epsilon_i)),
\Phi^{-1}(F_2(\xi_{ij}))\right),
\label{eq:gaussian_rho}
\end{equation}
where $\epsilon_i$ denotes the latent component of the binary outcome equation and $\xi_{ij}$ is the normalized latent utility difference in the polychotomous choice model.

The estimation procedure in Section \ref{sec:estimation} consistently recovers both the causal effects of alternatives on the binary outcome ($\tau_j$) and the transformed cross-equation gaussianized covariances $\rho_j$. Our central question, however, concerns the structural interpretation of these dependence parameters: how the covariance between the Gaussianized shocks relates to the primitive covariances between the unobserved components of the binary outcome and the latent utilities in the choice model (i.e., $Cov(\epsilon, u_j)$ or $Corr(\epsilon, u_j)$ ).

This section proceeds in three steps. First, we relate the estimated Gaussian copula parameter $\rho_j$ to the structural covariance $\Cov(\epsilon_i,\xi_{ij})$. Second, focusing on $\xi_{ij^*}$ for the realized choice, we derive a linear mapping from $\Cov(\epsilon_i,\xi_{ij^*})$ to the primitive covariances $\Cov(\epsilon_i,u_{ij^*})$ under joint normality. Third, we implement a simulation-based calibration to conduct inference.

These three steps establish that the estimated copula parameter is directly informative about the primitive covariances: under the null hypothesis that all primitive covariances $\Cov(\epsilon,u_j)$ are zero, the copula parameter associated with the realized choice is exactly zero, so that any dependence detected by our procedure must originate from nonzero covariances between the unobservable components of the two decisions.

\subsection{From Gaussian Copula Dependence to Structural Covariance}

The parameter $\rho_j$ in \eqref{eq:gaussian_rho} is a Gaussian copula parameter: it governs rank-based dependence after marginal standardization. In general, however, $\rho_{j}$ does not equal the structural covariance
\[
\mathrm{Cov}(\epsilon,\xi_{j}).
\]
For notational simplicity, denote by $\rho$ the copula parameter associated with the realized alternative $j^*$. Under a Gaussian copula with parameter $\rho$, the joint distribution of $(\epsilon,\xi_{j^*})$ admits the representation
\begin{equation}
(\epsilon,\xi_{j^*})
=
\big(
F_1^{-1}(\Phi(Z_{1})),
F_2^{-1}(\Phi(Z_{2}))
\big),
\quad
(Z_{1},Z_{2})
\sim
N\!\left(
0,
\begin{pmatrix}
1 & \rho \\
\rho & 1
\end{pmatrix}
\right).
\label{eq:copula_rep}
\end{equation}
It follows that
\begin{equation}
\mathrm{Cov}(\epsilon,\xi_{j^*})
=
\mathrm{Cov}
\Big(
F_1^{-1}(\Phi(Z_{1})),
F_2^{-1}(\Phi(Z_{2}))
\Big) \equiv m(\rho),
\label{eq:mapping}
\end{equation}
which is generally a nonlinear function of $\rho$ and of the marginal distributions $F_1$ and $F_2$. Only in the special case where both marginals are standard normal does $\rho$ coincide with $\mathrm{Corr}(\epsilon,\xi_{j^*})$, and therefore with the covariance. In general, the copula parameter must be translated into structural covariance through the nonlinear mapping in \eqref{eq:mapping}.

Although $m(\cdot)$ does not admit a closed form, it is well behaved, and the structural covariance vanish together and share the same sign:
\[
\rho = 0
\quad\Longleftrightarrow\quad
\Cov(\epsilon,\xi_{j^*}) = 0 .
\]
Consequently, restrictions on $\Cov(\epsilon,\xi_{j^*})$ translate one-for-one into restrictions on the copula parameter that our procedure estimates. The next subsection links $\Cov(\epsilon,\xi_{j^*})$, in turn, to the primitive covariances $\Cov(\epsilon,u_j)$.

\subsection{Structural Interpretation under Joint Normality}

We now establish a structural interpretation of the covariance between 
$\epsilon$ and the normalized latent utility difference $\xi_{j^*}$. 
To obtain a tractable mapping, we impose joint normality of the latent shocks.

\begin{assumption}[Joint Normality]
\[
\begin{pmatrix}
\epsilon \\
y_{1} \\
\vdots \\
y_{J}
\end{pmatrix}
\sim
N(\mu,\Sigma),
\]
where $y_{j}=V_{j}+u_{j}$ and $\Sigma$ has finite second moments.
\end{assumption}

Because $V_{j}=x_{1}\beta_j+z_j\delta$ contains no endogenous component,
\[
\Cov(\epsilon,y_{j})=\Cov(\epsilon,u_{j}) .
\]

We apply a result of \citet{siegel1993surprising} and \citet{rinott1994covariance}, which shows that for multivariate normal variables the covariance between a component and an order statistic equals a probability-weighted average of primitive covariances. In particular,
\[
\Cov(\epsilon,y_{(2)})
=
\sum_{j=1}^{J}
\Cov(\epsilon,y_{j})\,
\P(y_{j}=y_{(2)}).
\]

Using $\xi_{j^*}=y_{(2)}-u_{j^*}$, we obtain
\begin{equation}
\Cov(\epsilon,\xi_{j^*})
=
\sum_{j=1, j\neq j^*}^{J}
\Cov(\epsilon,u_{j})\,\P(y_{j}=y_{(2)})
-
\Cov(\epsilon,u_{j^*}).
\label{eq:cov_structural_normal_compact}
\end{equation}

Equation \eqref{eq:cov_structural_normal_compact} shows that, under joint normality, 
$\Cov(\epsilon,\xi_{j^*})$ is a linear function of the primitive covariance vector
\[
\big(
\Cov(\epsilon,u_{1}),\dots,\Cov(\epsilon,u_{J})
\big),
\]
with weights determined by second-order selection probabilities.

\medskip

\noindent
The linear mapping in \eqref{eq:cov_structural_normal_compact} relies on joint normality of the latent shocks. Absent this assumption, the relationship between $\Cov(\epsilon,\xi_{j^*})$ and the primitive covariances $\Cov(\epsilon,u_{j})$ need not be linear. In particular, the law of total covariance implies the presence of an additional selection-induced term involving conditional expectations. We provide the general decomposition, which holds without distributional assumptions, in Appendix \ref{app:non_normal_cov}. This decomposition clarifies that the linear representation above is specific to the multivariate normal case.

The structural null of interest is
\begin{equation}
H_0:\ \Cov(\epsilon,u_{j})=0 \quad \forall j.
\label{eq:struct_null}
\end{equation}

Under joint normality, this implies
\begin{equation}
H_0:\ \Cov(\epsilon,\xi_{j^*})=0.
\label{eq:reduced_null}
\end{equation}

In fact, a stronger statement holds: under Assumption 1, $\Cov(\epsilon,u_j)=0$ for all $j$ implies that $\epsilon$ is independent of $(u_1,\dots,u_J)$, and therefore of $\xi_{j^*}$, which is a function of the latent utility shocks conditional on covariates. 
\begin{equation}
\underbrace{\Cov(\epsilon,u_j)=0\ \ \forall j}_{\text{structural null \eqref{eq:struct_null}}}
\;\Longrightarrow\;
\underbrace{\rho_{j^*}=0}_{\text{copula parameter}}
\;\Longleftrightarrow\;
\underbrace{\Cov(\epsilon,\xi_{j^*})=0}_{\text{reduced-form null \eqref{eq:reduced_null}}} .
\label{eq:implication_chain}
\end{equation}
The chain in \eqref{eq:implication_chain} is what renders our procedure informative about the primitive covariances. Reading it from right to left: if the estimated dependence is nonzero, then $\Cov(\epsilon,\xi_{j^*})\neq 0$, and by \eqref{eq:cov_structural_normal_compact} this can occur only if $\Cov(\epsilon,u_j)\neq 0$ for at least one alternative $j$. Moreover, the calibrated quantity $m(\rho_{j^*})$ has a direct structural interpretation: by \eqref{eq:cov_structural_normal_compact}, it estimates the probability-weighted contrast
\(
\sum_{j}\Cov(\epsilon,u_{j})\,\P(y_{j}=y_{(2)})-\Cov(\epsilon,u_{j^*}),
\)
so the magnitude reported by our test measures a specific linear combination of the primitive covariances rather than an abstract dependence index. Two qualifications apply. First, because the test targets an implication of \eqref{eq:struct_null}, configurations in which nonzero primitive covariances exactly offset one another in \eqref{eq:cov_structural_normal_compact} are observationally equivalent to the null; rejection is therefore evidence of dependence between the unobservables, whereas failure to reject should not be read as confirming their independence. Second, because $\xi_{j^*}$ is non-normal and generated nonlinearly, the mapping $m(\cdot)$ between $\rho_{j^*}$ and $\Cov(\epsilon,\xi_{j^*})$ does not admit a closed form. We therefore implement a simulation-based calibration to test the structural null.

    \subsection{Simulation-Based Calibration and Structural Testing}

To connect the copula dependence parameter $\rho_{j^*}$ to the structural covariance 
$\Cov(\epsilon,\xi_{j^*})$, we construct a simulation-based estimate of the nonlinear mapping
\[
m(\rho)
=
\Cov\!\left(
F_1^{-1}(\Phi(Z_{1})),
F_2^{-1}(\Phi(Z_{2}))
\right),
\]
where $(Z_{1},Z_{2})$ are jointly normal with correlation $\rho$ as in \eqref{eq:mapping}.

\paragraph{Step 1: Marginal Distributions.}
The marginal distribution $F_2$ of $\xi_{j^*}$ is obtained from the estimated multinomial probit model described in Section \ref{sec:estimation}. In particular, the estimated parameters $\hat\theta$ imply the distribution $F_2(\cdot \mid \hat\theta)$ for the normalized latent utility difference $\xi_{j^*}$. 

The marginal distribution $F_1$ of $\epsilon$ follows from the maintained specification of the outcome equation and is standard normal under probit normalization.

\paragraph{Step 2: Simulating the Copula Representation.}
For a given candidate value of $\rho$, we draw $S$ independent realizations
\[
(Z_{1s},Z_{2s})
\sim
N\!\left(
0,
\begin{pmatrix}
1 & \rho \\
\rho & 1
\end{pmatrix}
\right),
\quad s=1,\dots,S.
\]
We then construct simulated structural shocks via the inverse-marginal transformation:
\[
\epsilon_s^{(\rho)}
=
\hat F_1^{-1}(\Phi(Z_{1s})),
\qquad
\xi_{j^*,s}^{(\rho)}
=
\hat F_2^{-1}(\Phi(Z_{2s}) \mid \hat\theta).
\]

The underlying uniform draws are held fixed across evaluations at different values of $\rho$ (common random numbers), so that the simulated mapping inherits the smoothness and strict monotonicity of $m(\cdot)$.
\paragraph{Step 3: Simulated Covariance.}
The structural covariance implied by $\rho$ is approximated by the sample covariance of the simulated draws:
\begin{equation}
\hat m_S(\rho)
=
\frac{1}{S}
\sum_{s=1}^S
\left(
\epsilon_s^{(\rho)} - \bar{\epsilon}^{(\rho)}
\right)
\left(
\xi_{j^*,s}^{(\rho)} - \bar{\xi}_{j^*}^{(\rho)}
\right),
\label{eq:sim_covariance}
\end{equation}
where
\[
\bar{\epsilon}^{(\rho)}
=
\frac{1}{S}\sum_{s=1}^S \epsilon_s^{(\rho)},
\qquad
\bar{\xi}_{j^*}^{(\rho)}
=
\frac{1}{S}\sum_{s=1}^S \xi_{j^*,s}^{(\rho)}.
\]
By the law of large numbers, $\hat m_S(\rho)$ converges to $m(\rho)$ as $S \to \infty$. We choose $S$ sufficiently large that the simulation error in $\hat m_S$ is negligible relative to the sampling variation of $\hat\rho_{j^*}$, so that the randomness in the test statistic below is driven by estimation rather than simulation.

\paragraph{Step 4: Null Distribution and Structural Test.}
Let $\hat\rho_{j^*}$ denote the copula dependence parameter estimated in
Section~\ref{sec:estimation}. Using the simulation-based mapping constructed above, the structural covariance implied by the estimated model is obtained by evaluating
\[
\widehat{\Cov}(\epsilon,\xi_{j^*})
=
\hat m_S(\hat\rho_{j^*}).
\]
Our objective is to test the structural null hypothesis \eqref{eq:struct_null} through its implication \eqref{eq:reduced_null}, which, by \eqref{eq:implication_chain}, is equivalent to $\rho_{j^*}=0$. Because the mapping $m(\rho)$ is nonlinear and depends on the estimated marginal distributions, the sampling distribution of $\widehat{\Cov}(\epsilon,\xi_{j^*})$ is not available in closed form. We therefore approximate its distribution \emph{under the null} using a parametric bootstrap that imposes $\rho_{j^*}=0$ and re-estimates the model in each replication, thereby accounting for estimation uncertainty in both the first-step parameters $\hat\theta$ and the copula parameters. For each bootstrap replication $b=1,\dots,B$:
\begin{enumerate}
\item Holding the covariates $(x_i,x_{1i},z)$ fixed at their sample values, draw latent utility shocks $u_i^{(b)}$ from the estimated multinomial probit error distribution and, independently, outcome shocks $\epsilon_i^{(b)}\sim N(0,1)$, thereby imposing the null of no dependence between the unobservables. Generate choices $I_i^{(b)}$ from Equation (1) and outcomes $m_i^{(b)}$ from Equation (2) using the estimated parameters $(\hat\beta,\hat\delta,\hat\gamma,\hat\tau)$.
\item Apply the two-step estimation procedure of Section \ref{sec:estimation} to the simulated sample, obtaining $\hat\theta^{(b)}$ and $\hat\rho_{j^*}^{(b)}$.
\item Repeat Steps 2--3 with the replication-specific marginals $F_2(\cdot\mid\hat\theta^{(b)})$ to compute the bootstrap analogue
\[
\widehat{\Cov}^{(b)}(\epsilon,\xi_{j^*})
=
\hat m_S^{(b)}\big(\hat\rho_{j^*}^{(b)}\big).
\]
\end{enumerate}
The empirical distribution of
$\big\{\widehat{\Cov}^{(b)}(\epsilon,\xi_{j^*})\big\}_{b=1}^B$
is used to construct the bootstrap critical value $C_{\alpha}$ at significance level $\alpha$ as the $(1-\alpha)$ quantile of $\big|\widehat{\Cov}^{(b)}(\epsilon,\xi_{j^*})\big|$. We reject the structural null hypothesis of no dependence between the unobservables if
\[
\left|\widehat{\Cov}(\epsilon,\xi_{j^*})\right| > C_{\alpha}.
\]
Because each replication imposes independence between the unobservable components and re-estimates all model parameters, the resulting distribution reflects the sampling variability of the test statistic under the null, including the estimation error inherited from the first-step multinomial probit estimates. Note also that, since $\hat m_S(\cdot)$ is strictly monotone, the test reaches the same rejection decision as a test based directly on $\hat\rho_{j^*}$ with critical values from the same replications; reporting the statistic on the covariance scale is preferred because, by \eqref{eq:cov_structural_normal_compact}, its magnitude is interpretable as a probability-weighted contrast of the primitive covariances $\Cov(\epsilon,u_j)$. Thus, rejection indicates that the covariance between the unobservable component affecting the binary outcome and the unobservable components of the latent utilities is sufficiently far from zero relative to the sampling variation expected under independence.\footnote{We assess the testing procedure through a Monte Carlo simulation study; see Appendix~\ref{sec:monte-carlo} for the data generating process and results.}

 \section{Data}

 We study the relationship between marriage prospects and college major choice using data from the National Longitudinal Survey of Youth 1997 cohort (NLSY97). The NLSY97 follows a nationally representative sample of individuals born between 1980 and 1984 who were aged 12 to 17 at the first interview in 1997. The initial sample contains 8,984 respondents and has been surveyed annually from 1997 to 2013 and biennially thereafter.\footnote{Source: U.S. Bureau of Labor Statistics, \url{https://www.nlsinfo.org}.} The survey provides detailed information on education, labor market outcomes, and family formation, allowing us to observe both college major choices and subsequent marital outcomes.

We focus on survey participants who graduated from higher education institutions. Conditioning on college graduates allows us to observe realized major choices and early-career labor market outcomes. Although this restriction excludes individuals who did not complete college, our objective is to study the allocation of students across majors rather than the decision to attend or complete college. Accordingly, our analysis should be interpreted as examining major choice and subsequent outcomes conditional on college completion.

The original NLSY97 sample contains 8,984 respondents. Restricting the sample to college graduates with observed income and working hours between 2005 and 2011 yields a final sample of 1,236 individuals. Table~\ref{tab: sumNLYS} presents summary statistics for male and female respondents. The respondents are between 31 and 35 years old in the 2015 survey wave. Using this information, we construct a binary marriage outcome variable, Ever Married, which indicates whether the respondent had been married at least once by 2015.

\begin{table}[t]
\centering
\caption{Summary Statistics of College Graduates}
\label{tab: sumNLYS}

\scriptsize
\setlength{\tabcolsep}{5pt}

\begin{threeparttable}
\begin{tabular}{L{4.2cm}C{1.3cm}C{1.3cm}C{1.3cm}C{1.3cm}C{1.3cm}C{1.3cm}}
\toprule
& \multicolumn{2}{c}{Female (56\%)} 
& \multicolumn{2}{c}{Male (44\%)} 
& \multicolumn{2}{c}{Total} \\
\cmidrule(lr){2-3}\cmidrule(lr){4-5}\cmidrule(lr){6-7}
Variables & Mean & Std.\ Dev. & Mean & Std.\ Dev. & Mean & Std.\ Dev. \\
\midrule

\multicolumn{7}{l}{\textit{Demographic Characteristics}} \\

Ever Married            & 0.65 & 0.48 & 0.60 & 0.49 & 0.63 & 0.48 \\
Birth Year              & 2.64 & 1.35 & 2.69 & 1.43 & 2.66 & 1.39 \\
Two-Year Degree         & 0.13 & 0.34 & 0.13 & 0.33 & 0.13 & 0.33 \\
Grad School             & 0.01 & 0.10 & 0.02 & 0.13 & 0.01 & 0.11 \\

\addlinespace
\multicolumn{7}{l}{\textit{Location and Race}} \\

North Central           & 0.19 & 0.39 & 0.21 & 0.41 & 0.20 & 0.40 \\
South                   & 0.37 & 0.48 & 0.34 & 0.47 & 0.36 & 0.48 \\
West                    & 0.18 & 0.39 & 0.16 & 0.37 & 0.17 & 0.38 \\
Rural                   & 0.70 & 0.46 & 0.68 & 0.47 & 0.69 & 0.46 \\
Black                   & 0.22 & 0.42 & 0.16 & 0.37 & 0.19 & 0.40 \\
Hispanic                & 0.15 & 0.36 & 0.14 & 0.35 & 0.15 & 0.35 \\
White                   & 0.61 & 0.49 & 0.69 & 0.46 & 0.65 & 0.48 \\

\addlinespace
\multicolumn{7}{l}{\textit{Academic Performance}} \\

GPA Term 1              & 2.73 & 1.12 & 2.69 & 1.04 & 2.71 & 1.09 \\
GPA Term 2              & 2.66 & 1.14 & 2.65 & 0.99 & 2.65 & 1.07 \\
GPA Term 3              & 2.66 & 1.16 & 2.57 & 1.08 & 2.62 & 1.13 \\
Overall GPA             & 2.82 & 0.89 & 2.63 & 0.95 & 2.73 & 0.92 \\

\addlinespace
\multicolumn{7}{l}{\textit{Major Choice}} \\

Business                & 0.15 & 0.36 & 0.18 & 0.38 & 0.16 & 0.37 \\
Education               & 0.10 & 0.30 & 0.03 & 0.18 & 0.07 & 0.25 \\
Computer \& Eng.        & 0.02 & 0.16 & 0.13 & 0.34 & 0.07 & 0.26 \\
Health                  & 0.12 & 0.32 & 0.03 & 0.17 & 0.08 & 0.27 \\
Other                   & 0.62 & 0.49 & 0.63 & 0.48 & 0.62 & 0.49 \\

\addlinespace
\multicolumn{7}{l}{\textit{Labor Market Outcomes}} \\

Avg.\ Income (\$1,000)  & 26.32 & 15.93 & 31.29 & 18.11 & 28.53 & 17.11 \\
Avg.\ Working Hours     & 1657.52 & 632.07 & 1850.29 & 758.57 & 1743.46 & 697.65 \\

\midrule
Observations            & \multicolumn{2}{c}{685} & \multicolumn{2}{c}{551} & \multicolumn{2}{c}{1236} \\
\bottomrule
\end{tabular}

\begin{tablenotes}[flushleft]
\scriptsize
\item \textit{Notes:} This table reports summary statistics for the sample of college graduates from the NLSY97. The first two columns report the mean and standard deviation for female graduates, the next two columns report the corresponding statistics for male graduates, and the last two columns report the combined sample statistics. The variable ``Ever Married'' indicates whether the individual had ever been married by 2015. ``Birth Year'' is measured as years since 1980. ``Avg.\ Income'' reports average annual income over 2005--2011 in thousands of dollars. ``Avg.\ Working Hours'' reports average annual working hours over 2005--2011.
\end{tablenotes}
\end{threeparttable}
\end{table}

The proportion of female college graduates in our sample amounts to 56\%, which is quite close to the population ratio of 58\% according to the National Center for Education Statistics' Digest of Education Statistics (2012). However, the NLSY97 dataset exhibits a slightly higher representation of male participants compared to female participants, possibly contributing to the difference in graduate ratios. An interesting contrast emerges in terms of marital status among graduates, particularly between females and males. In the sample we analyze, around 65\% of female college graduates have been married at least once by 2015. In contrast, the marriage rate for male college graduates is lower by five percentage points.

Academic performance is measured using grade point averages from the first three college terms as well as overall GPA. These measures capture students' academic ability and are included as controls in the analysis of major choice. Prior research shows that students often revise their major expectations after receiving information about their academic performance in college \citep{stin_14}. Consistent with this literature, we observe that female students have slightly higher GPAs on average across the first three terms of college.\footnote{Standard pre-college measures such as SAT/ACT scores or high school GPA could provide additional ability controls, but these variables are not consistently available in the NLSY97 transcript data.}

College majors are coded using the 2010 College Course Map classification available in the NLSY97 transcript data. When respondents have multiple college transcripts, we assign the major associated with the largest number of completed credits. Following this classification, we group detailed fields of study into four broad categories: Business and related fields, Health-related programs, Education, and Engineering and Computer Science. These fields represent some of the most common majors among college graduates and exhibit substantial gender differences in enrollment patterns. The distribution of major choices among college graduates in our sample is reported in Table~\ref{tab: sumNLYS}. Reassuringly, the distribution of majors in our sample closely resembles that observed in the broader population of U.S. college graduates, suggesting that the sample provides a reasonable representation of national patterns in field-of-study choice.

 As shown in Table~\ref{tab: sumNLYS}, major choices differ markedly by gender. Male graduates are substantially more likely to select Engineering and Computer Science fields, whereas Education and Health majors are disproportionately chosen by women. For example, 13\% of men in the sample graduate with degrees in Engineering or Computer Science compared to only 2\% of women, while 10\% of women major in Education compared to 3\% of men. Although the precise shares differ somewhat from national statistics, the overall gender patterns are consistent with those documented in the broader literature \citep{bronson}. The distribution of majors within the NLSY97 also remains relatively stable over time, as shown in Figure~\ref{fig:pop_majors} in Appendix~\ref{app:additional-tables-figures}.

\begin{table}[htbp]
\centering
\caption{Marriage Expectations of College Candidates and Marital Status of College Graduates}
\label{tab: mar_exp_real}

\scriptsize
\setlength{\tabcolsep}{6pt}

\begin{threeparttable}

\begin{tabular}{L{4cm}C{2.5cm}C{2.5cm}C{2.5cm}}
\toprule

& \multicolumn{2}{c}{Stated Marriage Expectation (\%)} 
& Marital Status \\

\cmidrule(lr){2-3}\cmidrule(lr){4-4}

& 2000 & 2001 & 2015 \\

\midrule
\multicolumn{4}{l}{\textit{By Major}} \\

Education          & 52.15 & 58.24 & 0.75 \\
Business           & 38.80 & 49.57 & 0.62 \\
Computer \& Eng.   & 44.53 & 41.53 & 0.66 \\
Health             & 41.42 & 50.53 & 0.63 \\
Other              & 43.65 & 45.44 & 0.62 \\

\addlinespace

Total              & 43.37 & 47.03 & 0.63 \\

\addlinespace
\multicolumn{4}{l}{\textit{By Gender}} \\

Female             & 47.12 & 51.11 & 0.65 \\
Male               & 38.72 & 41.66 & 0.60 \\

\midrule

Observations       & 1181 & 570 & 1236 \\

\bottomrule
\end{tabular}

\begin{tablenotes}[flushleft]
\scriptsize
\item \textit{Notes:} Stated marriage expectation measures represent the self-assessed probability (in percent) that respondents expect to get married within the next five years. Marital status is an indicator variable equal to one if the individual has ever been married by 2015.
\end{tablenotes}

\end{threeparttable}
\end{table}

The NLSY97 also contains information on respondents' expectations regarding marriage. Table~\ref{tab: mar_exp_real} reports average self-stated marriage expectations and realized marital status by major and gender. In survey waves 2000 and 2001, respondents were asked to report the probability (in percent) that they expected to marry within the next five years.\footnote{The survey also collects information on cohabitation, but we focus on legal marriage outcomes.} These expectations vary across both gender and major fields. For example, in 2000 female respondents reported a 47\% probability of marrying within the next five years compared with 39\% among males. Education majors report the highest expected marriage probabilities, while Business and Engineering majors report lower expectations.

Actual marital outcomes exhibit similar patterns. By 2015, approximately 75\% of Education majors had been married at least once, compared with marriage rates between 62\% and 66\% for the other major categories. The correlation between marriage expectations and realized marital outcomes, together with the variation across majors and genders, suggests that marriage prospects may be related to students' major choices.

Finally, the sequence of events in our data is as follows. Students enroll in college and accumulate grade information during the first three terms, which we use as measures of academic performance. During the early survey waves (2000–2001) before deciding their major, respondents are questioned to report expectations regarding marriage within the next five years. Students subsequently complete a major, which we observe from transcript data.  After graduation, we observe labor market outcomes between 2005 and 2011, including earnings and working hours. Finally, we observe realized marital status by 2015, when respondents are between 31 and 35 years old.

\subsection{Descriptive Regressions}

Next, we provide some descriptive results with two sets of regressions where the outcomes are (i) college major choice and (ii) whether or not a person has been married by 2015. These descriptive regression outcomes serve as a natural reference point for comparison against the estimates derived from our proposed methodology, which we elaborate on in the subsequent sections. Table~\ref{tab: MNL_reduced} offers an overview of the descriptive estimation outcomes for college major choices. To account for monetary motivations in major selection, we incorporate counterfactual normalized hourly earnings predictions that are generated using both sample data and additional exogenous variations.\footnote{The earnings estimation results are available in Table \ref{tab:earnings}. The computation of counterfactual wages includes degree types, college GPA (which uses students’ initial GPA as their forecast for their final GPA), and average working hours as supplementary exogenous variations. The computation of counterfactual working hours includes attendance at a second college or not as an additional exogenous variations. Estimation results are presented in Table \ref{tab:workinghours}. Details of expected earnings and annual working hours regressions are available in Appendix \ref{app:earning and working hours}.} We normalize the coefficient of expected normalized earnings to 1; therefore, we can interpret the results in terms of normalized earnings. Among the observed factors, gender emerges as the most influential and substantial determinant of major choices. Specifically, female students exhibit a statistically significant and quantitatively substantial inclination against selecting Computer Science and Engineering programs. Conversely, they are more inclined to opt for Education and Health programs. Notably, performance measures play a significant role in influencing major choices. Particularly, within the first three terms of college education, students' GPA has discernible effects on their choice of major. For instance, higher GPA scores in the second semester correspond to a greater tendency to choose Computer Science and Engineering programs. Similarly, students with elevated GPAs in the third semester demonstrate an increased likelihood of choosing Education programs.

\begin{table}[htbp]
\centering
\caption{Descriptive Regressions: Major Choice}
\label{tab: MNL_reduced}

\scriptsize
\setlength{\tabcolsep}{5pt}

\begin{threeparttable}

\begin{tabular}{L{2.5cm}C{1.2cm}C{1.2cm}C{1.2cm}C{1.2cm}C{1.2cm}C{1.2cm}C{1.2cm}C{1.2cm}}
\toprule
& \multicolumn{2}{c}{Education}
& \multicolumn{2}{c}{Business}
& \multicolumn{2}{c}{Health}
& \multicolumn{2}{c}{Comp.\ \& Eng.} \\
\cmidrule(lr){2-3}\cmidrule(lr){4-5}\cmidrule(lr){6-7}\cmidrule(lr){8-9}

& Est. & Std.\ Err.
& Est. & Std.\ Err.
& Est. & Std.\ Err.
& Est. & Std.\ Err. \\
\midrule

Female & 1.273 & 0.281 & -0.264 & 0.161 & 0.883 & 0.280 & -1.965 & 0.283 \\

GPA Term 1 & -0.122 & 0.119 & -0.025 & 0.085 & 0.106 & 0.121 & 0.104 & 0.134 \\

GPA Term 2 & 0.150 & 0.139 & -0.032 & 0.091 & -0.114 & 0.123 & 0.343 & 0.159 \\

GPA Term 3 & 0.272 & 0.139 & 0.037 & 0.084 & 0.091 & 0.117 & -0.164 & 0.126 \\

\addlinespace

Expected Earnings & \multicolumn{2}{c}{Yes} & \multicolumn{2}{c}{Yes} & \multicolumn{2}{c}{Yes} & \multicolumn{2}{c}{Yes} \\

Location Controls & \multicolumn{2}{c}{Yes} & \multicolumn{2}{c}{Yes} & \multicolumn{2}{c}{Yes} & \multicolumn{2}{c}{Yes} \\

Race Controls & \multicolumn{2}{c}{Yes} & \multicolumn{2}{c}{Yes} & \multicolumn{2}{c}{Yes} & \multicolumn{2}{c}{Yes} \\

Age Controls & \multicolumn{2}{c}{Yes} & \multicolumn{2}{c}{Yes} & \multicolumn{2}{c}{Yes} & \multicolumn{2}{c}{Yes} \\

\bottomrule
\end{tabular}

\begin{tablenotes}[flushleft]
\scriptsize
\item \textit{Notes:} This table reports multinomial logit regression estimates for students' major choices. The coefficient on normalized expected earnings is fixed at one. Remaining majors are grouped as ``Other'' and serve as the base category. Marginal effects corresponding to these coefficients are reported in the top panel of Table~\ref{tab: MNP_structural_reduced_Mar}.
\end{tablenotes}

\end{threeparttable}
\end{table}

Table~\ref{tab: Mar_reduced} presents the descriptive regression results of marital status determinants under three different covariate sets. It's important to note that, in all three specifications, the constant term is omitted from the regressions in order to present coefficients for all possible college major choices, including other majors. Among the significant findings, a noticeable difference in marital status emerges between graduates from the Education major and graduates from other majors. Those who majored in Education exhibit higher marriage rates than graduates from other majors, a trend observed across all three specifications. Furthermore, the descriptive analysis also underscores the disparity between females and males as well as the impact of average earnings on observed marital status.\footnote{Additional descriptive regression results that segment college major choices and marriage outcomes for male and female graduates are provided in Tables \ref{tab: MNL_reduced_gender} and \ref{tab:results22} in the Appendix. These outcomes elucidate the divergences between male and female students in terms of their major preferences and marital outcomes during their thirties. The disparities identified between male and female students' college major choices and marital statuses underscore the significance of unobservable factors, motivating us to carry out separate analyses for male and female cohorts.}

\begin{table}[htbp]
\centering
\caption{Descriptive Regressions: Marriage Outcome}
\label{tab: Mar_reduced}

\scriptsize
\setlength{\tabcolsep}{5pt}

\begin{threeparttable}

\begin{tabular}{L{5cm}C{1.2cm}C{1.2cm}C{1.2cm}C{1.2cm}C{1.2cm}C{1.2cm}}
\toprule
& \multicolumn{2}{c}{(1)} 
& \multicolumn{2}{c}{(2)} 
& \multicolumn{2}{c}{(3)} 
\\
\cmidrule(lr){2-3}\cmidrule(lr){4-5}\cmidrule(lr){6-7}

& Est. & Std.\ Err. 
& Est. & Std.\ Err. 
& Est. & Std.\ Err. \\

\midrule

Female & 0.168 & 0.079 & 0.194 & 0.079 & 0.186 & 0.080  \\

Education & 1.178 & 0.192 & 0.921 & 0.210 & 1.013 & 0.228 \\

Business & 0.852 & 0.139 & 0.560 & 0.169 & 0.646 & 0.188 \\

Health & 0.929 & 0.179 & 0.634 & 0.204 & 0.715 & 0.218 \\

Com.\ \& Eng. & 0.877 & 0.168 & 0.541 & 0.201 & 0.618 & 0.214 \\

Other & 0.849 & 0.116 & 0.584 & 0.145 & 0.668 & 0.166 \\

\addlinespace

Avg.\ Observed Earnings 
&  &  & 0.007 & 0.002 & 0.009 & 0.003  \\

Avg.\ Observed Working Hours$^{*}$ 
&  &  &  &  & -0.066 & 0.063 \\

\addlinespace

Location Controls 
& \multicolumn{2}{c}{Yes} 
& \multicolumn{2}{c}{Yes} 
& \multicolumn{2}{c}{Yes}  \\

Race Controls 
& \multicolumn{2}{c}{Yes} 
& \multicolumn{2}{c}{Yes} 
& \multicolumn{2}{c}{Yes}  \\

Age Controls 
& \multicolumn{2}{c}{Yes} 
& \multicolumn{2}{c}{Yes} 
& \multicolumn{2}{c}{Yes}  \\

\midrule

Observations 
& \multicolumn{2}{c}{1236} 
& \multicolumn{2}{c}{1236} 
& \multicolumn{2}{c}{1236} \\

\bottomrule
\end{tabular}

\begin{tablenotes}[flushleft]
\scriptsize
\item \textit{Notes:} This table reports probit regression estimates for realized marriage outcomes under four alternative covariate specifications. Average observed earnings and working hours are computed using post-graduation labor market outcomes. $^{*}$Average working hours are divided by 1000 in the regressions. Marginal effects corresponding to these estimates are reported in columns (1)–(3) of Table~\ref{tab: marginaleffects}.
\end{tablenotes}

\end{threeparttable}
\end{table}

\section{Testing the Effects of Marriage Prospects on College Major Choice}

In this section we apply the estimation framework developed above to examine whether marriage prospects influence college major choice. The model allows an unobserved marriage prospect factor to affect both major selection and subsequent marital outcomes. Identification relies on the dependence between the unobservable components of these two decisions. In the copula-based specification, this dependence is captured by the vector of correlation parameters $\boldsymbol{\rho}$. If the unobserved determinants of major choice are statistically correlated with those affecting marital outcomes, this indicates the presence of latent factors influencing both decisions. Within our framework, these common factors are interpreted as marriage prospects that shape students' major choices.

\subsection{Empirical Specification}

\noindent \textbf{College Major Choice.} We focus on undergraduate major decisions and abstract from graduate education and college duration.\footnote{Less than 5\% of individuals in our sample obtain graduate degrees.} Student $i$ derives utility $V_{ij}$ from graduating with major $j$, which depends on labor market expectations and other major-specific attributes.

Labor market expectations consist of expected earnings and working hours. Let $w_{ij}$ denote student $i$'s expected earnings in major $j$, which depend on individual characteristics $x_i$. Working hours, denoted by $h_{ij}$, capture differences in employment intensity and work flexibility across fields. To measure the financial return to each major independent of labor supply differences, we define expected hourly earnings as
\begin{equation}
v_{ij}^{l}=  \E\left[ \frac{w_{ij}}{h_{ij}} \mid x_{1i} \right],
\end{equation}%
where $v_{ij}^{l}$ represents the expected hourly earnings of student $i$ choosing major $j$, conditional on student characteristics.

The remaining utility components capture non-pecuniary aspects of major choice, such as personal preferences, marriage prospects, and major-specific attributes. We express these as:
\begin{equation}
v_{ij}^{n}=x_{1i}\beta _{j}+u_{ij},
\end{equation}%
where $x_{1i}$ includes student characteristics (e.g., gender, academic performance, race, and regional factors), $\beta_{j}$ represents major-specific parameters, and $u_{ij}$ denotes unobserved preferences, including marriage expectations and leisure preferences. We assume $u_{ij}$ is independent of observed characteristics, following standard choice modeling assumptions.

Students select the major that maximizes expected utility:
\begin{equation}
j_i^*= \underset{j \in \{1, \hdots, J\}}{\arg \max} \ V_{ij},
\end{equation}%
where $V_{ij} = v_{ij}^l+ v_{ij}^n =  \E\left[ \frac{w_{ij}}{h_{ij}} \mid x_{1i} \right] + x_{1i} \beta_j + u_{ij}$.

\smallskip

\noindent \textbf{Marital Outcome.} We model the marriage probability as a function of observable characteristics, college major, and unobserved factors:
\begin{equation}
m_{i} = \mathbf{1}\Big(x_i \mathbf{\gamma} + \sum_{j=1}^J j_i\tau_j + \epsilon_{i} >0 \Big),
\end{equation}%
where $m_i$ is a marriage indicator (equal to 1 if married at least once, 0 otherwise). Here, $x_i$ includes graduate $i$'s observable attributes (e.g., age, gender, race, and geographic location), alongside actual earnings and working hours—distinct from expected earnings and hours used in the major choice model. This differentiation controls for post-graduation effects on marriage outcomes.

Identification of the dependence parameters relies on an exclusion restriction: students' GPAs from their first three college terms enter the major choice equation but are excluded from the  marriage outcome equation.\footnote{Following the identification discussion in Section \ref{sec:2}, early college GPA is included in $x_{1}$ but not in $x$.} These early performance measures are realized before the major decision and generate substantial variation in choice probabilities across alternatives, as documented in Tables \ref{tab: MNL_reduced} and
\ref{tab: MNP_structural}. The identifying assumption is that, conditional on the covariates in the outcome equation---including realized post-graduation earnings and working hours---early college grades do not directly affect marital outcomes measured more than a decade later.\footnote{ This assumption is not innocuous: early grades may partly reflect persistent traits, such as conscientiousness, that could influence family formation directly. Two considerations mitigate this concern. First, the marriage equation controls for realized earnings and working hours, which absorb the principal channels through which ability or effort would plausibly affect marital outcomes. Second, if the exclusion restriction failed because grades proxy such traits, the resulting spurious dependence would persist even after conditioning on stated marriage expectations; instead, as shown in Section \ref{sec:validation}, the estimated dependence disappears once the stated marriage expectation measure is included, a pattern that is difficult to reconcile with a violation of this form. For marital outcomes, we control for observed annual earnings and working hours, which enter the marriage equation but not the major choice equation, since only expected---rather than realized---labor market outcomes can affect the major decision. This approach leverages variation in the data and mitigates reliance on functional form assumptions for identification.}

\subsection{Results}

Table \ref{tab: selection_eq} presents the estimation results for the marriage outcome equation alongside the vector of correlation coefficients derived from our structural estimation approach and Table \ref{tab: MNP_structural} presents the estimation results for the major choice component. We employ the simulated maximum likelihood estimation (SMLE), setting the number of simulations to 250.

\begin{table}[htbp]
\centering
\caption{Structural Estimation Results: Marriage Outcome}
\label{tab: selection_eq}
\scriptsize
\setlength{\tabcolsep}{6pt}

\begin{tabular}{L{5cm}C{2.5cm}C{2.5cm}}
\toprule
& \multicolumn{2}{c}{(1)}  \\
\cmidrule(lr){2-3}
 & Estimate & Std. Err.  \\
\midrule

Female                  & -0.006 & 0.085  \\
Education               & 1.231  & 0.772  \\
Business                & -0.656 & 0.354  \\
Health                  & 0.562  & 0.650 \\
Com.\ \& Eng.           & 0.005  & 0.509  \\
Other                   & 0.692  & 0.664 \\

Avg.\ Observed Earnings & -0.003 & 0.002\\
Avg.\ Observed Hours    & 0.011  & 0.051 \\

\addlinespace
$\rho_1$ & 0.029 & 0.294  \\
$\rho_2$ & 0.877 & 0.092 \\
$\rho_3$ & 0.280 & 0.262  \\
$\rho_4$ & 0.355 & 0.181  \\
$\rho_5$ & 0.547 & 0.661  \\

\midrule
Location Controls & \multicolumn{2}{c}{Yes}  \\
Race Controls     & \multicolumn{2}{c}{Yes}  \\
Age Controls      & \multicolumn{2}{c}{Yes}  \\

\addlinespace
Simulated Correlation &  0.451 & 0.023  \\
Bootstrap Critical Value $C_{0.05}$ &  \multicolumn{2}{c}{0.046}   \\
\bottomrule
\end{tabular}
\begin{tablenotes}
\scriptsize
\item \textit{Notes:} This table presents estimation results of the marriage outcome equation and the vector correlation coefficients of the proposed estimation for the whole sample. To obtain choice probabilities, we take the number of GHK simulations as 250. Standard errors are calculated from the outer product of the log-likelihoods gradient. $\boldsymbol{\rho}$ parameters are estimated with a transformation function $2(2/(1+exp(\rho)) -1)$. Estimation results for major choice component of the joint estimation are presented separately in Table \ref{tab: MNP_structural}. 
\end{tablenotes}
\end{table}

Our primary focus is on the dependence parameters $\boldsymbol{\rho}$, which capture the dependence between the unobservable components across the two equations. As reported in Table~\ref{tab: selection_eq}, several elements of $\boldsymbol{\rho}$---most notably $\rho_2$ and $\rho_4$---are statistically different from zero, suggesting the presence of nontrivial dependence between the unobserved determinants of marital outcomes and college major choices. Because the copula parameters do not directly measure the structural dependence of interest, we implement the simulation-based calibration test described in Section~\ref{testing}. Using the estimated marginal distributions of the error terms, we map the estimated copula parameter into the implied dependence between the unobservable component of the marriage equation and the normalized latent utility difference arising from the major choice model, $\hat m_S(\hat\rho_{j^*})$. The resulting simulated correlation is $0.451$, which far exceeds the bootstrap critical value of $C_{0.05}=0.046$ obtained from 1,000 replications that impose independence between the unobservable components and re-estimate the model in each replication. We therefore reject the null hypothesis of no dependence between the unobservables at the 5\% level. This finding provides empirical support for the presence of systematic dependence between the two decision processes and is consistent with the hypothesis that marriage prospects play a role in shaping individuals’ college major choices.

\begin{table}[htbp]
\centering
\caption{Structural Estimation Results: Major Choice}
\label{tab: MNP_structural}

\scriptsize
\setlength{\tabcolsep}{5pt}

\begin{threeparttable}

\begin{tabular}{L{2.5cm}C{1.2cm}C{1.2cm}C{1.2cm}C{1.2cm}C{1.2cm}C{1.2cm}C{1.2cm}C{1.2cm}}
\toprule
& \multicolumn{2}{c}{Education}
& \multicolumn{2}{c}{Business}
& \multicolumn{2}{c}{Health}
& \multicolumn{2}{c}{Comp.\ \& Eng.} \\
\cmidrule(lr){2-3}\cmidrule(lr){4-5}\cmidrule(lr){6-7}\cmidrule(lr){8-9}

& Est. & Std.\ Err.
& Est. & Std.\ Err.
& Est.& Std.\ Err.
& Est. & Std.\ Err. \\
\midrule

Female & 0.936 & 0.109 & -0.347 & 0.068 & 0.394 & 0.097 & -1.051 & 0.094 \\

GPA Term 1 & -0.105 & 0.050 & 0.034 & 0.036 & 0.036 & 0.046 & 0.106 & 0.052 \\

GPA Term 2 & 0.048 & 0.056 & -0.106 & 0.039 & 0.050 & 0.050 & 0.014 & 0.057 \\

GPA Term 3 & 0.120 & 0.054 & 0.036 & 0.044 & -0.015 & 0.052 & -0.018 & 0.062 \\

\addlinespace

Expected Earnings
& \multicolumn{2}{c}{Yes}
& \multicolumn{2}{c}{Yes}
& \multicolumn{2}{c}{Yes}
& \multicolumn{2}{c}{Yes} \\

Location Controls
& \multicolumn{2}{c}{Yes}
& \multicolumn{2}{c}{Yes}
& \multicolumn{2}{c}{Yes}
& \multicolumn{2}{c}{Yes} \\

Race Controls
& \multicolumn{2}{c}{Yes}
& \multicolumn{2}{c}{Yes}
& \multicolumn{2}{c}{Yes}
& \multicolumn{2}{c}{Yes} \\

Age Controls
& \multicolumn{2}{c}{Yes}
& \multicolumn{2}{c}{Yes}
& \multicolumn{2}{c}{Yes}
& \multicolumn{2}{c}{Yes} \\

\bottomrule
\end{tabular}

\begin{tablenotes}[flushleft]
\scriptsize
\item \textit{Notes:} This table reports the college major choice component of the structural estimation results for the full sample. Results for the marriage outcome component of the joint model are reported in Table \ref{tab: selection_eq}. Average marginal effects corresponding to these estimates are reported in Table~\ref{tab: MNP_structural_reduced_Mar}.
\end{tablenotes}

\end{threeparttable}
\end{table}

To compare the descriptive and structural estimates reported in Tables \ref{tab: Mar_reduced} and \ref{tab: selection_eq}, we report marginal effects in Table \ref{tab: marginaleffects}, since probit coefficients are not directly comparable across specifications. The descriptive regressions in Table \ref{tab: Mar_reduced} (column (3)) indicate that graduating with an Education major increases the probability of marriage by about 12 percentage points (pp.) relative to other majors, while the marginal effects for Business, Health, Computer Science \& Engineering, and other majors are broadly similar. In contrast, the structural estimates reveal larger and more heterogeneous effects. Graduating with an Education major increases the probability of marriage by approximately 19 pp., whereas graduating with a Business major decreases it by about 50 pp. These differences between reduced-form and structural marginal effects suggest the presence of omitted-variable bias in specifications that do not account for correlations between unobservable factors.

\begin{table}[htbp]
\centering
\caption{Marginal Effects Comparisons: Marriage Outcome}
\label{tab: marginaleffects}

\scriptsize
\setlength{\tabcolsep}{5pt}

\begin{threeparttable}

\begin{tabular}{L{3cm}C{1.2cm}C{1.2cm}C{1.2cm}C{1.2cm}C{1.2cm}C{1.2cm}C{1.2cm}C{1.2cm}}
\toprule
& \multicolumn{6}{c}{Table \ref{tab: Mar_reduced}} 
& \multicolumn{2}{c}{Table \ref{tab: selection_eq}} \\
\cmidrule(lr){2-7}\cmidrule(lr){8-9}

& \multicolumn{2}{c}{(1)}
& \multicolumn{2}{c}{(2)}
& \multicolumn{2}{c}{(3)}
& \multicolumn{2}{c}{(1)} \\
\cmidrule(lr){2-3}\cmidrule(lr){4-5}\cmidrule(lr){6-7}\cmidrule(lr){8-9}

& AME & Std.\ Err.
& AME & Std.\ Err.
& AME & Std.\ Err.
& AME & Std.\ Err. \\

\midrule

Education       & 0.427 & 0.068 & 0.349 & 0.075 & 0.389 & 0.081 & 0.439 & 0.216 \\
Business        & 0.294 & 0.048 & 0.206 & 0.060 & 0.243 & 0.067 & -0.257 & 0.053 \\
Health          & 0.305 & 0.064 & 0.215 & 0.074 & 0.251 & 0.079 & 0.192 & 0.212 \\
Comp.\ \& Eng.  & 0.327 & 0.060 & 0.224 & 0.072 & 0.258 & 0.077 & -0.024 & 0.162 \\
Other           & 0.309 & 0.040 & 0.228 & 0.052 & 0.265 & 0.059 & 0.246 & 0.250 \\

\bottomrule
\end{tabular}

\begin{tablenotes}[flushleft]
\scriptsize
\item \textit{Notes:} This table reports average marginal effects of college major on the probability of marriage. Columns (1)–(3) correspond to the reduced-form probit estimates reported in Table~\ref{tab: Mar_reduced}, while the final column reports results based on the structural estimation presented in Table~\ref{tab: selection_eq}. Standard errors are computed using a nonparametric bootstrap with 1,000 replications.
\end{tablenotes}

\end{threeparttable}
\end{table}

In Table \ref{tab: MNP_structural_reduced_Mar}, we present the marginal effect estimates derived from both the descriptive regressions (top panel) and  the proposed structural estimation method (bottom panel) for college major choice. By examining the marginal effects, we can compare the magnitude of the impact of various variables on major choices and provide clear interpretations of our empirical analysis. The marginal effect estimates indicate the change in the probability of choosing a specific college major compared to choosing majors that are not specifically analyzed in our study (i.e., Others). For example, the marginal effect estimate for female students shows that the probability of choosing an Education field is over 8 pp. higher than the probability of choosing other majors.

The disparities in marginal effect estimates underscore the variations in results that the descriptive regressions fail to capture. Notably, there are significant differences in the effects of expected earnings and gender on choice probabilities. For instance, female college students are more likely to choose Education fields and less likely to choose Business and Computer \& Engineering fields compared to Health and other fields, with these effects ranging between 7 and 10 pp. in the structural estimates. Conversely, the marginal effects obtained from the reduced-form model suggest suggest that female students are slightly more likely to choose the Business field, while their probability of choosing Computer \& Engineering fields being more than 48\%.

\subsection{Validation exercise}
\label{sec:validation}

Because our testing procedure relies on the correlation between unobservable components of the econometric model, the results provide only suggestive evidence regarding the role of marriage prospects in major choice. To complement this approach, we conduct a validation exercise using a direct proxy for marriage prospects.

Specifically, we use survey responses from the NLSY97 that record individuals’ self-assessed probability of getting married within the next five years (the stated marriage expectation variable). Since these responses were reported prior to students’ college major decisions, they provide exogenous variation that can be used to directly examine the role of marriage prospects in major choice. We therefore construct a stated marriage expectation measure and incorporate it into our empirical framework. This validation exercise is feasible in our setting due to the availability of this unique survey measure, which is rarely observed in other applications.

We implement the proposed testing procedure after adding the stated marriage expectation variable to the existing set of covariates.\footnote{Descriptive regression results examining the relationship between the stated marriage expectation variable and major choice are reported in Table \ref{tab: MNL_reduced1} in Appendix \ref{sec:appendixvalidation}.} 
Our structural approach allows for unobservable factors to jointly influence both marital outcomes and major choice, a feature that cannot be accommodated in the reduced-form MNL regressions reported in Table \ref{tab: MNL_reduced1}. By including the stated marriage expectation variable in the structural model, we are able to isolate the effects of marriage prospects directly while allowing the remaining correlation parameters to capture other potential sources of unobserved dependence.

\begin{table}[t]
\centering
\caption{Structural Estimation Results: Marriage Outcome}
\label{tab: corr_eq}

\scriptsize
\setlength{\tabcolsep}{6pt}

\begin{threeparttable}

\begin{tabular}{L{4.2cm}C{1.75cm}C{1.75cm}C{1.75cm}C{1.75cm}}
\toprule
& \multicolumn{2}{c}{Coefficients} & \multicolumn{2}{c}{Marginal Effects} \\
\cmidrule(lr){2-3}\cmidrule(lr){4-5}
 & Estimate & Std.\ Err. & AME & Std.\ Err. \\
\midrule

Female                  & 0.135 & 0.092 & 0.129 & 0.087 \\
Education               & 0.696 & 1.100 & 0.176 & 0.262 \\
Business                & 0.442 & 1.022 & 0.129 & 0.268 \\
Health                  & 0.501 & 0.786 & 0.150 & 0.222 \\
Comp.\ \& Eng.          & 0.376 & 0.596 & 0.109 & 0.179 \\
Other                   & 0.445 & 0.419 & 0.151 & 0.129 \\
Avg.\ Observed Earnings & 0.012 & 0.003 & 0.003 & 0.001 \\
Avg.\ Observed Working Hours & -0.082 & 0.069 & -0.025 & 0.021 \\
\textbf{Marriage Exp.}  & 0.006 & 0.001 & 0.002 & 0.001 \\

\addlinespace
$\rho_1$ & -0.008 & 0.385 &  &  \\
$\rho_2$ & 0.010  & 0.584 &  &  \\
$\rho_3$ & 0.012  & 0.288 &  &  \\
$\rho_4$ & -0.002 & 0.222 &  &  \\
$\rho_5$ & -0.001 & 0.503 &  &  \\

\midrule

Location Controls & \multicolumn{4}{c}{Yes} \\
Race Controls     & \multicolumn{4}{c}{Yes} \\
Age Controls      & \multicolumn{4}{c}{Yes} \\

\addlinespace
Simulated Correlation & 0.001 & 0.029    \\
Bootstrap Critical Value $C_{0.05}$ &  \multicolumn{2}{c}{0.051}  \\
\bottomrule
\end{tabular}

\begin{tablenotes}[flushleft]
\scriptsize
\item \textit{Notes:} This table reports structural estimation results for the marriage outcome equation and the dependence parameters for the full sample. Choice probabilities are computed using 250 GHK simulations. Standard errors are calculated from the outer product of the log-likelihood gradient. The correlation parameters $\boldsymbol{\rho}$ are estimated using the transformation function $2(2/(1+exp(\rho)) -1)$ Estimation results for the major choice component of the joint model are reported in Table~\ref{tab: MNP_me}
\end{tablenotes}

\end{threeparttable}
\end{table}

Table \ref{tab: corr_eq} reports the estimated coefficients and marginal effects for the marriage outcome equation, along with the dependence parameters obtained from the proposed testing procedure when the stated marriage expectation variable is included in the model. The results indicate that once marriage expectations are explicitly controlled for, none of the estimated correlation parameters are statistically different from zero. This contrast becomes clear when comparing Tables \ref{tab: selection_eq} and \ref{tab: corr_eq}, where the former reported significant dependence between the unobservable components of the two equations.

Consistent with this finding, the simulated correlation is $0.001$, well below the bootstrap critical value of $C_{0.05}=0.051$, so we fail to reject the null hypothesis of no dependence between the unobservable components of the marriage and major choice equations. Taken together, these results suggest that the previously estimated dependence across equations largely disappears once marriage expectations are incorporated directly into the model. This evidence supports the interpretation that marriage prospects are the primary source of the unobserved dependence identified in the baseline specification.

Table \ref{tab: MNP_me_marginalchoice} reports the average marginal effects of selected variables on major choice probabilities from the structural model. The marginal effects represent the change in the probability of choosing a given major relative to other majors. The results indicate that expected earnings play a central role in major choice. An increase in the expected earnings associated with a major significantly raises the probability of selecting that major. For example, a \$1000 increase in normalized expected earnings increases the probability of choosing an Education major by more than 7 pp., and the effect exceeds 30 pp. for Computer Science and Engineering majors.

Gender differences are also substantial. Female students are approximately 50\% less likely to select Computer Science and Engineering majors relative to male students, holding other factors constant. In addition, early academic performance has a strong influence on major choice, with GPA in the first three terms significantly shifting choice probabilities. Finally, marriage expectations also affect major choice, although their magnitudes are smaller. For instance, a 10\% increase in the self-assessed probability of getting married within the next five years increases the likelihood of choosing an Education major by approximately 0.4 pp.

\section{Heterogeneous Effects}

The significant effects of the female indicator in both the structural estimation results (Tables \ref{tab: MNP_structural} and \ref{tab: MNP_me}) and the descriptive regressions (Tables \ref{tab: MNL_reduced_gender} and \ref{tab:results22}) reveal substantial differences between male and female students in both college major choice and marital outcomes in their 30s. These disparities suggest differences in decision-making processes that may also alter the role of unobservable factors across genders. To investigate these differences, we estimate the model separately for male and female participants.

We incorporate the stated marriage expectation measure directly into the econometric framework to obtain clearer estimates.\footnote{Estimation results from our proposed method without the stated marriage expectation measure are reported in Appendix \ref{app:additional-tables-figures} and indicate significant impacts of unobservable factors.} Tables \ref{tab: selection_eq_gender} and \ref{tab: MNPstructuralgender} present the estimation results for the marriage outcome equation, including the vector of correlation coefficients, and the parameters of the college major choice component, respectively. The results continue to confirm the main finding: higher stated marriage expectations significantly increase the likelihood of marriage for both male and female graduates. Moreover, once the stated marriage expectation measure is included, the estimated dependence parameters are small and statistically insignificant for both groups, and the simulated correlations---$0.031$ for females and $0.004$ for males---fall below their respective bootstrap critical values ($C_{0.05}=0.049$ and $C_{0.05}=0.051$), so the test no longer detects dependence between the unobservable components for either gender.

However, notable differences emerge between male and female graduates. First, the effects of selected college majors on marriage outcomes are not statistically significant for male graduates. In contrast, for female graduates, Education and Health majors are associated with significantly higher probabilities of marriage in their 30s. Table \ref{tab: selection_eq_gender} reports the corresponding marginal effects that quantify these relationships.

To examine how marriage prospects affect major choice, Table \ref{tab: MNP_me_marginalchoice_gender} reports marginal effects from the structural model separately for female and male students. The results show that the relative importance of expected earnings, early GPA, and the self-assessed probability of marriage remains similar across genders, although their magnitudes differ.

For female students, a \$1000 increase in normalized expected earnings for Education majors increases the probability of choosing that major by more than 6\%, while the effect exceeds 47\% for Computer Science and Engineering majors. In contrast, for male students the same increase in expected earnings for Education majors does not significantly affect choice probabilities. Similar to female students, the largest variation in male choice probabilities arises from increases in normalized expected earnings for Computer Science and Engineering majors; a \$1000 increase raises the probability of choosing this major by more than 11\%.

\begin{table}[htbp]
\centering
\caption{Structural Estimation Results by Gender: Marriage Outcome}
\label{tab: selection_eq_gender}

\scriptsize
\setlength{\tabcolsep}{5pt}

\begin{threeparttable}

\begin{tabular}{L{4cm}C{1cm}C{1.2cm}C{1cm}C{1.2cm}C{1cm}C{1.2cm}C{1cm}C{1.2cm}}
\toprule
& \multicolumn{4}{c}{Female} & \multicolumn{4}{c}{Male} \\
\cmidrule(lr){2-5}\cmidrule(lr){6-9}
& \multicolumn{2}{c}{Coefficients} & \multicolumn{2}{c}{Marginal Effects}
& \multicolumn{2}{c}{Coefficients} & \multicolumn{2}{c}{Marginal Effects} \\
\cmidrule(lr){2-3}\cmidrule(lr){4-5}\cmidrule(lr){6-7}\cmidrule(lr){8-9}
& Est.& Std.\ Err. & AME & Std.\ Err.
& Est. & Std.\ Err. & AME & Std.\ Err. \\
\midrule

Education & 1.294 & 0.594 & 0.152 & 0.163 & 0.165 & 0.872 & 0.012 & 0.142 \\
Business & 1.099 & 1.030 & 0.117 & 0.194 & -0.057 & 0.857 & -0.017 & 0.151 \\
Health & 0.939 & 0.459 & 0.121 & 0.129 & 0.471 & 0.954 & 0.047 & 0.155 \\
Comp.\ \& Eng. & 1.186 & 0.889 & 0.134 & 0.176 & 0.016 & 0.830 & -0.008 & 0.149 \\
Other & 0.964 & 0.401 & 0.118 & 0.143 & 0.093 & 0.520 & 0.005 & 0.093 \\
Avg.\ Observed Earnings & 0.072 & 0.110 & 0.001 & 0.006 & 0.035 & 0.110 & 0.001 & 0.007 \\
Avg.\ Observed Working Hours & -0.222 & 0.112 & -0.025 & 0.029 & 0.111 & 0.095 & 0.014 & 0.019 \\
\textbf{Marriage Exp.} & 0.005 & 0.002 & 0.001 & 0.001 & 0.011 & 0.003 & 0.001 & 0.001 \\

\addlinespace

$\rho_{1}$ & 0.016 & 0.130 &  &  & -0.002 & 0.164 &  &  \\
$\rho_{2}$ & 0.043 & 0.446 &  &  & 0.004 & 0.383 &  &  \\
$\rho_{3}$ & 0.026 & 0.122 &  &  & 0.005 & 0.196 &  &  \\
$\rho_{4}$ & 0.036 & 0.220 &  &  & 0.024 & 0.573 &  &  \\
$\rho_{5}$ & 0.041 & 0.465 &  &  & 0.002 & 0.189 &  &  \\

\midrule

Location Controls
& \multicolumn{2}{c}{Yes} & \multicolumn{2}{c}{Yes}
& \multicolumn{2}{c}{Yes} & \multicolumn{2}{c}{Yes} \\

Race Controls
& \multicolumn{2}{c}{Yes} & \multicolumn{2}{c}{Yes}
& \multicolumn{2}{c}{Yes} & \multicolumn{2}{c}{Yes} \\

Age Controls
& \multicolumn{2}{c}{Yes} & \multicolumn{2}{c}{Yes}
& \multicolumn{2}{c}{Yes} & \multicolumn{2}{c}{Yes} \\

\addlinespace
Simulated Correlation & 0.031 & 0.039  & &  & 0.004 & 0.042  \\
Bootstrap Critical Value $C_{0.05}$ &  \multicolumn{2}{c}{0.049}  & & &  \multicolumn{2}{c}{0.051}\\
\bottomrule
\end{tabular}

\begin{tablenotes}[flushleft]
\scriptsize
\item \textit{Notes:} This table reports structural estimation results for the marriage outcome equation and the dependence parameters estimated separately for female and male graduates. Choice probabilities are computed using 250 GHK simulations. Standard errors are obtained via bootstrap simulations. The correlation parameters $\boldsymbol{\rho}$ are estimated using the  transformation function $2(2/(1+exp(\rho)) -1)$.
Estimation results for the major choice component of the joint model are reported in Table~\ref{tab: MNPstructuralgender}.
\end{tablenotes}

\end{threeparttable}
\end{table}

Early academic performance also influences major choice, although its effects differ across genders. The largest differences arise for Education and Computer Science and Engineering majors. For male students, a higher first-term GPA increases the probability of choosing Computer Science and Engineering, whereas for female students it decreases that probability. Conversely, a higher first-term GPA raises the probability of selecting an Education major for female students but does not significantly affect male students’ choices.

Marriage expectations also affect major choice differently across genders. For female students, a higher self-assessed probability of getting married within the next five years increases the probability of choosing Education and Business majors. For male students, these effects are generally not statistically significant. For both genders, however, higher marriage expectations significantly reduce the likelihood of selecting Computer Science and Engineering majors. The magnitude of this effect differs considerably: for female students, a 10\% increase in the self-assessed probability of marriage increases the probability of choosing an Education major by almost 2 pp. and decreases the probability of choosing a Computer Science and Engineering major by more than 15 pp. For male students, the same increase leads to a statistically insignificant change in Education major choice and a decrease of roughly 0.5 pp. in the probability of selecting Computer Science and Engineering.

Finally, the effects of average observed earnings and annual working hours on marital status are important for controlling socioeconomic differences and revealing gender heterogeneity. Tables \ref{tab:FM_nome} and \ref{tab: selection_eq_gender} show that annual working hours affect females and males in opposite directions. Differences in leisure preferences or household labor division may explain this pattern. If traditional gender roles influence marriage decisions in the data, males may be expected to work more to support the household, while females may devote more time to housework and childcare, leaving less time for market work, consistent with the empirical findings.

Although the proposed method may not provide highly detailed structural interpretations, it demonstrates its usefulness in generating empirical evidence that helps researchers investigate the role of unobservable factors and interpret results appropriately when such factors potentially play an important role.

\section{Conclusion}

This paper develops a new econometric framework for analyzing the effects of an unobservable factor—a confounding variable—that jointly influences a polychotomous choice (unordered multinomial choice) and a subsequent binary outcome. Our main methodological contribution is a formal test for the presence of such common unobservable factors. The framework extends binary endogenous choice models to settings in which the endogenous explanatory variable is polychotomous, and it models the joint dependence between the unobserved components of the choice and outcome equations through a copula-based specification. This structure separates the identification of the causal effects of the polychotomous choice on the outcome from the dependence induced by unobserved confounding factors. By exploiting the ordering structure of latent utilities in the choice model, we construct a test for whether the unobserved factors affecting the binary outcome are systematically related to the latent utilities that determine choice among alternatives. Because the normalization of latent utilities implies that only transformed versions of the unobservable components are identified, we further develop a simulation-based calibration procedure that maps the estimated copula dependence parameter into the implied covariance between the structural shocks. As such, the proposed method also serves as a diagnostic tool for confounding unobservable factors and a complementary approach to dynamic choice models, offering an initial test for the presence of unobserved dependence across decision stages.

We apply the method to examine the role of marriage prospects in college major choice. Our analysis highlights substantial differences in marital outcomes among college graduates and incorporates an unobserved marriage prospect factor within the joint econometric framework. Consistent with prior research \citep{keane_wolphin_97, arcidiocono_04, stin_14}, we find that expected earnings, early academic performance, and gender play central roles in determining college major selection. Beyond these well-established determinants, our results indicate that marriage prospects significantly influence major choice decisions, supporting evidence in the literature \citep{beffy_12, wiswall_zafar_15, wiswall2021human} that non-pecuniary considerations play an important role in educational choices. Because many relevant factors—such as personality traits, life expectations, and educational preferences—are rarely observed in available datasets, the proposed method offers researchers a useful tool for investigating the role of unobservable non-pecuniary factors in similar contexts.

The finding that marriage prospects affect college major choice also has implications for labor market policy. Students’ educational decisions directly shape the distribution of skills in the workforce, and factors related to marriage and family expectations can influence these decisions, particularly for women. Despite rising college enrollment and graduation rates among women, their labor market participation remains lower than that of men. Policies that support family-friendly work environments may therefore play an important role in shaping educational and career choices. By improving workplace flexibility and family support policies, employers and policymakers can encourage broader participation in certain career paths, increase the labor force participation of college-educated women, and contribute to a more balanced distribution of skills in the labor market.

\bibliographystyle{apalike}
\footnotesize
\bibliography{sample}
 
\newpage 

\appendix
\clearpage
\setcounter{page}{1}

\numberwithin{equation}{section}
\numberwithin{table}{section}
\numberwithin{figure}{section}

\section{Estimation Steps}
\label{section:appendixEst}

\subsection{Multinomial Choice}

\noindent \textbf{Multinomial Probit - GHK Implementation.} Specifically, the vector of $\mathbf{u}_{J}$ follows a joint distribution with mean $\mu_{u}= (0, \hdots, 0)^{\prime}$ and covariance

\[
\Sigma_{\mathbf{u}} = 
\begin{pmatrix} 
 1 &   &a  \\ 
\vdots &  \ddots & \\ 
 a &  & 1
\end{pmatrix}_{(J \times J)}
\]

A lower triangular matrix, denoted as $h_J$, exists in a way that it satisfies the equation $h_J \cdot h_j^{\prime} = \Sigma_{\mathbf{u}}$. Subsequently, by considering $\mathbf{u}_{J} =h_J\mathbf{\eta}_{J}$, where $\mathbf{\eta}_{J} = (\eta_1, . . . , \eta_J)^{\prime}$ follows a standard $J$-variate distribution. Given that $h_J$ is a lower triangular matrix, a recursive formula can be established as follows:

\begin{align*}
    u_1 =&  h_{11} \eta_1 \\
    u_2 =&  h_{11} \eta_1 + h_{22}\eta_2 \\
    & \vdots \\
    u_J =&  h_{11} \eta_1 + \hdots + h_{JJ}\eta_J \\
\end{align*}
where $h_{ij} = h(i,j)$. Then, we can make simulated draws and approximate choice probabilities for each major and agent. Let's denote  $p_{i,j}(\theta)$ as the likelihood that student $i$'s choice is major $j$ and can be approximated by the following procedure:

\begin{itemize}
\item Step 1: Draw $\eta$ from $N(0,1)$ and calculate $1-(\Phi(\frac{(-x_{1i}  \beta_{j^*} - z_{j^*}\delta)}{h_{11}}| \eta_1)$.

\item Step 2:  Draw $\eta$ from $N(0,1)$ truncated at $\frac{(-x_{1i}  \beta_{j^*} - z_{j^*}\delta - h_{11} \eta_1 )}{h_{22}}$ and calculate $1- \Phi(\frac{(-x_{1i}  \beta_{j^*} - z_{j^*}\delta- h_{11} \eta_1 )}{h_{22}}| \eta_1,\eta_2)$.
\begin{center}
 $\vdots$   
\end{center}
\item Step $J$-1: Draw $\eta$ from $N(0,1)$ truncated at $\frac{(-x_{1i}  \beta_{j^*} - z_{j^*}\delta - h_{11} \eta_1 - \hdots - h_{(J-1),(J-1)}\eta_{(J-1)} )}{h_{(J-1),(J-1)}}$ and calculate

$1- \Phi(\frac{(-x_{1i}  \beta_{j^*} - z_{j^*}\delta- h_{11} \eta_1 -  \hdots - h_{(J-1),(J-1)}\eta_{(J-1)} )}{h_{(J-1),(J-1)}}| \eta_1,\eta_2, \eta_{(J-2)})$.
\end{itemize}

\bigskip

\noindent \textbf{Multinomial Logit.} We can also use a multinomial logistic model (MNL). MNL is the most common specification because of its analytical solution for the probabilities. Under the MNL model, we can write $F_2 (\cdot)$ as
\[
F_2(x_{1i}  \beta_{j^*} +z_{j^*}\delta )= \frac{e^{x_{1i}  \beta_{j^*} +z_{j^*}\delta}}{1+ \sum_{k=1}^{J-1} e^{x_{1i}  \beta_{k} +z_{k}\delta }}
\]

\subsection{Derivation of Bivariate Gaussian Copula}

We can write $\P(m_i=0, I_i=j | x_i, z_i) $ using the equality in Equation \ref{prob_0j} and the bivariate normal distribution.
\begin{align*}
C(F_1(-x_i\gamma & -\tau_j)  , F_2(x_{1i}  \beta_{j^*} +z_{j^*}\delta) = B (\Phi^{-1}(F_1(-x_i\gamma -\tau_j)), \Phi^{-1}(F_2(V_{ij}))) \\
=& B( \epsilon_i \leq -x_i\gamma -\tau_j, \xi_{ij} \leq \Phi^{-1}(F_2(V_{ij}))) | x, z) \\
=& \Phi( \epsilon_i \leq -x_i\gamma -\tau_j | x, z) \; \Phi( \xi_{ij} \leq \Phi^{-1}(F_2(V_{ij}))) | \epsilon_i \leq -x_i\gamma -\tau_j, x, z) \\
=&  \Phi(-x_i\gamma -\tau_j) \; \frac{1}{1- \Phi(-x_i\gamma- \tau_j)} \int_{- \infty}^{(-x_i\gamma- \tau_j)} \Phi [ (\Phi^{-1}(F_2(V_{ij}))) + \rho_j \epsilon_i)/ (1-\rho_j^2)^{1/2}] \phi(\epsilon) d \epsilon \\
=& \int_{- \infty}^{(-x_i\gamma- \tau_j)} \Phi [ (\Phi^{-1}(F_2(V_{ij}))) + \rho_j \epsilon_i)/ (1-\rho_j^2)^{1/2}] \phi(\epsilon) d \epsilon 
\end{align*}

\section{General Covariance Decomposition without Joint Normality}
\label{app:non_normal_cov}

This appendix derives a general relationship between 
$\Cov(\epsilon_i,\xi_{ij^*})$ and the primitive covariances 
$\Cov(\epsilon_i,u_{ij})$ that holds without distributional assumptions.

Recall that the normalized latent utility difference is
\[
\xi_{ij^*} = y_{i(2)} - u_{ij^*},
\]
where $y_{i(2)}$ denotes the second order statistic of the latent utilities
$y_{ij}=V_{ij}+u_{ij}$.

By linearity of covariance,
\begin{equation}
\Cov(\epsilon_i,\xi_{ij^*})
=
\Cov(\epsilon_i,y_{i(2)})
-
\Cov(\epsilon_i,u_{ij^*}).
\label{eq:cov_general_appendix}
\end{equation}

To characterize $\Cov(\epsilon_i,y_{i(2)})$, apply the law of total covariance:
\[
\Cov(X,Y)
=
\E[\Cov(X,Y\mid Z)]
+
\Cov(\E[X\mid Z],\E[Y\mid Z]).
\]

Let $Z=(u_{i1},\dots,u_{iJ})$. Then
\begin{align}
\Cov(\epsilon_i,y_{i(2)})
&=
\E\!\big[\Cov(\epsilon_i,y_{i(2)} \mid Z)\big]
+
\Cov\!\big(
\E[\epsilon_i\mid Z],
\E[y_{i(2)}\mid Z]
\big).
\label{eq:ltc_step}
\end{align}

Conditional on $Z$, the second order statistic equals one of the latent utilities, so
\[
y_{i(2)}
=
\sum_{j=1}^J y_{ij}\,\mathbf{1}\{y_{ij}=y_{i(2)}\}.
\]
Using this representation,
\[
\E\!\big[\Cov(\epsilon_i,y_{i(2)} \mid Z)\big]
=
\sum_{j=1}^{J}
\Cov(\epsilon_i,u_{ij})\,\P(y_{ij}=y_{i(2)}).
\]

Combining with \eqref{eq:cov_general_appendix}, we obtain
\begin{align}
\Cov(\epsilon_i,\xi_{ij^*})
&=
\sum_{j=1}^{J}
\Cov(\epsilon_i,u_{ij})\,\P(y_{ij}=y_{i(2)})
\nonumber\\
&\quad
+
\Cov\!\big(
\E[\epsilon_i\mid Z],
\E[y_{i(2)}\mid Z]
\big)
-
\Cov(\epsilon_i,u_{ij^*}).
\label{eq:cov_total_appendix}
\end{align}

Equation \eqref{eq:cov_total_appendix} holds without distributional assumptions. 
It shows that, in general, $\Cov(\epsilon_i,\xi_{ij^*})$ depends not only on the primitive covariances $\Cov(\epsilon_i,u_{ij})$, weighted by second-order selection probabilities, but also on an additional term involving conditional expectations.

If $\epsilon_i$ is independent of $(u_{i1},\dots,u_{iJ})$, then both
$\Cov(\epsilon_i,u_{ij})=0$ for all $j$ and the conditional expectation term vanishes, implying
\[
\Cov(\epsilon_i,\xi_{ij^*})=0.
\]

Under joint normality, the conditional expectation term reduces to a linear function of the primitive covariances, yielding the linear representation presented in the main text.

\section{Expected Earnings and Working Hours Calculations} \label{app:earning and working hours}

To capture the effect of the expected labor market conditions on major choice, we predict average earnings and annual working hours of college graduates from different majors. Beyond major-related variations, we include student characteristics ($z$) to encompass other contributing factors and achieve diverse predictions. Our linear regression model for generating counterfactual earnings and working hours is outlined as follows:

\[
\text{Labor market outcomes}= x_{1i}\delta + \sum_{j=1}^J j_i\kappa_j + \tilde{x}_{1i}\lambda + u_i
\]
where $z$ and $j$ denote student characteristics and major dummies, respectively. It is important to note that, in contrast to student characteristics related to major choice ($x_1$), our regression specifications involve additional variables $(\tilde{x}_1)$ that are likely independent of students' major choices. The inclusion of these supplementary variables allows us to identify labor market conditions without relying on assumptions related to functional forms.

\begin{table}[htbp]
\centering
\caption{Linear Regression Results of Earnings}
\label{tab:earnings}

\scriptsize
\setlength{\tabcolsep}{6pt}

\begin{threeparttable}

\begin{tabular}{L{6.5cm}C{2cm}C{2cm}}
\toprule
Dependent Variable & \multicolumn{2}{c}{Earnings} \\
\cmidrule(lr){2-3}
 & Estimate & Std.\ Err. \\
\midrule

Female & -2.47 & 0.85 \\
Birth Year & -2.32 & 0.29 \\
Hispanic & -1.64 & 1.22 \\
Black & -3.13 & 1.14 \\
Two-Year Degree & -2.02 & 1.22 \\
Graduate School & -4.82 & 3.56 \\
Rural & 1.38 & 0.91 \\
West & 1.70 & 1.30 \\
South & 1.83 & 1.07 \\
North East & 3.82 & 1.19 \\
Overall GPA & 0.02 & 0.00 \\
Education & 6.52 & 2.72 \\
Business & 11.36 & 2.36 \\
Health & 13.64 & 2.64 \\
Comp.\ \& Eng. & 17.02 & 2.64 \\
Other Majors & 9.03 & 2.20 \\
Working Hours & 0.01 & 0.00 \\

\bottomrule
\end{tabular}

\begin{tablenotes}[flushleft]
\scriptsize
\item \textit{Notes:} This table reports linear regression estimates of the effects of observable characteristics and chosen college majors on graduates' earnings. These estimates are used to construct counterfactual earnings for college graduates in subsequent analysis.
\end{tablenotes}

\end{threeparttable}
\end{table}

\begin{table}[htbp]
\centering
\caption{Linear Regression Results of Annual Working Hours}
\label{tab:workinghours}

\scriptsize
\setlength{\tabcolsep}{6pt}

\begin{threeparttable}

\begin{tabular}{L{6.5cm}C{2cm}C{2cm}}
\toprule
Dependent Variable & \multicolumn{2}{c}{Annual Working Hours} \\
\cmidrule(lr){2-3}
 & Estimate & Std.\ Err. \\
\midrule

Female & -175.62 & 40.46 \\
Birth Year & -82.63 & 13.96 \\
Second College & -21.92 & 48.92 \\
Hispanic & 20.77 & 58.06 \\
Black & 42.02 & 53.26 \\
Two-Year Degree & 25.35 & 58.83 \\
Graduate School & 242.25 & 167.22 \\
Rural & -60.74 & 43.56 \\
West & -39.58 & 61.90 \\
South & 45.17 & 51.43 \\
North East & 166.03 & 57.23 \\
Education & 2120.40 & 95.92 \\
Business & 2123.80 & 75.27 \\
Health & 2037.74 & 95.07 \\
Comp.\ \& Eng. & 2088.60 & 91.22 \\
Other Majors & 2007.92 & 64.55 \\

\bottomrule
\end{tabular}

\begin{tablenotes}[flushleft]
\scriptsize
\item \textit{Notes:} This table reports linear regression estimates of the effects of observable characteristics and chosen college majors on graduates' annual working hours. These estimates are used to construct counterfactual annual working hours for college graduates in subsequent analysis.
\end{tablenotes}

\end{threeparttable}
\end{table}

Tables \ref{tab:earnings} and \ref{tab:workinghours} display the estimation results utilized in constructing the counterfactual earnings and annual working hours of college graduates. In the earnings regression, we introduce degree type as an additional variable, positing that students may have preferences for education duration that are independent of students' major choice. Furthermore, we incorporate working hours to account for earnings variations based on the number of hours worked. In the working hours regression, we include a second college variable that indicates whether the terminal college degree was obtained by a second college attendee or not.

\section{Monte Carlo Simulations}
\label{sec:monte-carlo}

To examine the performance of the proposed testing procedure, we
conduct a Monte Carlo simulation study. This appendix describes the data
generating process, the parameter values, the correlation structures we
consider, and the resulting size and power of the test.

\subsection{Data Generating Process}

In order to match with our application, we select the sample size as 1,236 as in Table \ref{tab: sumNLYS}.

\paragraph{Polychotomous choice.}
The polychotomous choice is generated from three explanatory variables. The
variable $x_{11}$ is binary and equal to one with probability $0.5$; the
variables $x_{12}$ and $x_{13}$ are continuous, drawn independently from a
uniform distribution on $[1.8, 4]$. Following our empirical application, we
consider $J = 5$ alternatives and normalize the coefficients of the last
alternative to zero. Each coefficient vector $\beta_k = (\beta_{k1}, \dots,
\beta_{k5})'$ collects the effect of regressor $x_{1k}$ across the five
alternatives:
\begin{align*}
    \beta_1 &= (0.717,\ -0.239,\ 0.485,\ -1.452,\ 0), \\
    \beta_2 &= (-0.104,\ -0.007,\ 0.046,\ 0.064,\ 0), \\
    \beta_3 &= (0.113,\ -0.013,\ -0.052,\ 0.198,\ 0).
\end{align*}

\paragraph{Binary outcome.}
The binary outcome depends on the chosen alternative and on two additional
regressors: a binary variable $x_{1}$ equal to one with probability $0.5$, and a
variable that does not enter the polychotomous choice, drawn from a normal
distribution with mean zero and standard deviation $2$. The associated
coefficients are
\begin{align*}
    \gamma &= (0.134,\ -0.086), \\
    \tau   &= (0.445,\ 0.695,\ 0.441,\ 0.501,\ 0.376),
\end{align*}
where $\gamma$ collects the coefficients on the two binary-equation regressors
and $\tau$ collects the effect of each chosen alternative on the binary outcome.

\subsection{Correlation Structure}

The error terms of the five latent utilities and of the binary outcome are drawn
jointly from a multivariate normal distribution with covariance matrix
\[
\Sigma =
\begin{pmatrix}
    1 & 0 & 0 & 0 & 0 & \rho_1 \\
    0 & 1 & 0 & 0 & 0 & \rho_2 \\
    0 & 0 & 1 & 0 & 0 & \rho_3 \\
    0 & 0 & 0 & 1 & 0 & \rho_4 \\
    0 & 0 & 0 & 0 & 1 & \rho_5 \\
    \rho_1 & \rho_2 & \rho_3 & \rho_4 & \rho_5 & 1
\end{pmatrix},
\]
where $\rho = (\rho_1, \rho_2, \rho_3, \rho_4, \rho_5)$ collects the correlations
between the error term of the binary outcome and the error terms of the five
latent-utility alternatives. The latent-utility errors are mutually independent,
so the only source of dependence between the two equations is $\rho$.

\subsection{Design}

We assess the test under one null design and two alternative designs:
\begin{itemize}
    \item[--] \textbf{Size (no correlation):} $\rho = (0, 0, 0, 0, 0)$.
    \item[--] \textbf{Power, design (i):} $\rho = (0.3,\ 0,\ -0.2,\ 0,\ 0.5)$.
    \item[--] \textbf{Power, design (ii):} $\rho = (0.5,\ 0,\ -0.4,\ 0,\ 0.7)$.
\end{itemize}

\subsection{Results}

Under the null design, the estimated correlations recover the true zeros:
$\tilde{\rho} = (0.059,\ -0.009,\ -0.046,$ $ -0.031,\ 0.006)$ with standard errors
$(0.164,\ 0.387,\ 0.256,\ 0.163,\ 0.382)$. The calibration test statistic has
mean $-0.004$ and standard error $0.003$, well below the bootstrap critical
value of $0.049$; the procedure therefore fails to reject and correctly
indicates the absence of correlation.

Under both alternative designs the test rejects the null. The calibration test
statistics of $0.2218$ (design (i)) and $0.1403$ (design (ii)) exceed the
bootstrap critical value of $0.049$, so the procedure correctly detects the
presence of correlation. Table~\ref{tab:mc-results} summarizes the results.

\begin{table}[htbp]
\centering
\scriptsize
\caption{Monte Carlo results for the calibration test.}
\label{tab:mc-results}
\begin{tabular}{lccc}
\hline\hline
Design & $\rho$ & Test statistic (s.e.) & Decision \\
\hline
Null       & $(0,0,0,0,0)$          & $-0.004\ (0.003)$ & Fail to reject \\
Power (i)  & $(0.3,0,-0.2,0,0.5)$   & $0.2218\ (0.028)$ & Reject \\
Power (ii) & $(0.5,0,-0.4,0,0.7)$   & $0.1403\ (0.029)$ & Reject \\
\hline
\multicolumn{4}{l}{\scriptsize Bootstrap critical value: $0.049$.} \\
\hline\hline
\end{tabular}
\end{table}

\section{Additional Tables and Figures} \label{app:additional-tables-figures}

\subsection{Average Treatment Effect: Major Choice}
\label{sec:appendixmajorchoicemarginal}

\begin{table}[htbp]
\centering
\caption{Marginal Effects Comparisons: Major Choice}
\label{tab: MNP_structural_reduced_Mar}

\scriptsize
\setlength{\tabcolsep}{5pt}

\begin{threeparttable}

\begin{tabular}{L{2.5cm}C{1.2cm}C{1.2cm}C{1.2cm}C{1.2cm}C{1.2cm}C{1.2cm}C{1.2cm}C{1.2cm}}
\toprule
& \multicolumn{2}{c}{Education}
& \multicolumn{2}{c}{Business}
& \multicolumn{2}{c}{Health}
& \multicolumn{2}{c}{Comp.\ \& Eng.} \\
\cmidrule(lr){2-3}\cmidrule(lr){4-5}\cmidrule(lr){6-7}\cmidrule(lr){8-9}

& AME & Std.\ Err.
& AME & Std.\ Err.
& AME & Std.\ Err.
& AME & Std.\ Err. \\

\midrule

\multicolumn{9}{l}{\textit{Panel A: Descriptive Estimates}} \\

Female & 0.0212 & 0.0039 & 0.0192 & 0.0010 & 0.0000 & 0.0000 & -0.4831 & 0.0172 \\
GPA Term 1 & -0.0062 & 0.0010 & -0.0029 & 0.0002 & 0.0000 & 0.0000 & 0.0391 & 0.0017 \\
GPA Term 2 & 0.0053 & 0.0009 & -0.0104 & 0.0004 & 0.0000 & 0.0000 & 0.1241 & 0.0055 \\
GPA Term 3 & 0.0169 & 0.0017 & 0.0045 & 0.0003 & 0.0000 & 0.0000 & -0.0045 & 0.0006 \\
Expected Earnings & 0.0474 & 0.0076 & 0.0708 & 0.0037 & 0.0000 & 0.0000 & 0.3560 & 0.0156 \\

\addlinespace
\multicolumn{9}{l}{\textit{Panel B: Structural Estimates}} \\

Female & 0.0842 & 0.0094 & -0.0982 & 0.0048 & 0.0000 & 0.0000 & -0.0746 & 0.0134 \\
GPA Term 1 & -0.0179 & 0.0017 & 0.0072 & 0.0004 & 0.0000 & 0.0000 & 0.0123 & 0.0017 \\
GPA Term 2 & 0.0117 & 0.0011 & -0.0264 & 0.0007 & 0.0000 & 0.0000 & 0.0065 & 0.0010 \\
GPA Term 3 & 0.0169 & 0.0017 & 0.0045 & 0.0003 & 0.0000 & 0.0000 & -0.0045 & 0.0006 \\
Expected Earnings & 0.1531 & 0.0152 & 0.2251 & 0.0061 & 0.0001 & 0.0000 & 0.1273 & 0.0181 \\

\midrule

Location Controls
& \multicolumn{2}{c}{Yes}
& \multicolumn{2}{c}{Yes}
& \multicolumn{2}{c}{Yes}
& \multicolumn{2}{c}{Yes} \\

Race Controls
& \multicolumn{2}{c}{Yes}
& \multicolumn{2}{c}{Yes}
& \multicolumn{2}{c}{Yes}
& \multicolumn{2}{c}{Yes} \\

Age Controls
& \multicolumn{2}{c}{Yes}
& \multicolumn{2}{c}{Yes}
& \multicolumn{2}{c}{Yes}
& \multicolumn{2}{c}{Yes} \\

\bottomrule
\end{tabular}

\begin{tablenotes}[flushleft]
\scriptsize
\item \textit{Notes:} This table reports average marginal effects for college major choice obtained from the descriptive multinomial logit estimates in Table~\ref{tab: MNL_reduced} and the structural estimates in Table~\ref{tab: MNP_structural}. Standard errors are computed using a nonparametric bootstrap with 1,000 replications.
\end{tablenotes}

\end{threeparttable}
\end{table}

\begin{table}[htbp]
\centering
\caption{Marginal Effects from Structural Estimation: Major Choice}
\label{tab: MNP_me_marginalchoice}

\scriptsize
\setlength{\tabcolsep}{5pt}

\begin{threeparttable}

\begin{tabular}{L{2.5cm}C{1.2cm}C{1.2cm}C{1.2cm}C{1.2cm}C{1.2cm}C{1.2cm}C{1.2cm}C{1.2cm}}
\toprule
& \multicolumn{2}{c}{Education}
& \multicolumn{2}{c}{Business}
& \multicolumn{2}{c}{Health}
& \multicolumn{2}{c}{Comp.\ \& Eng.} \\
\cmidrule(lr){2-3}\cmidrule(lr){4-5}\cmidrule(lr){6-7}\cmidrule(lr){8-9}

& AME & Std.\ Err.
& AME & Std.\ Err.
& AME & Std.\ Err.
& AME & Std.\ Err. \\
\midrule

\textbf{Marriage Exp.} & 0.00043 & 0.00006 & -0.00022 & 0.00001 & 0.00000 & 0.00001 & 0.00007 & 0.00002 \\

Female & 0.02565 & 0.00364 & 0.01920 & 0.00101 & 0.00000 & 0.00001 & -0.54290 & 0.00561 \\

GPA Term 1 & -0.00938 & 0.00120 & -0.00150 & 0.00008 & 0.00000 & 0.00001 & 0.02397 & 0.00089 \\

GPA Term 2 & 0.00415 & 0.00058 & -0.00732 & 0.00032 & 0.00000 & 0.00001 & 0.05990 & 0.00169 \\

GPA Term 3 & 0.01150 & 0.00148 & 0.00265 & 0.00013 & 0.00000 & 0.00001 & -0.03116 & 0.00115 \\

Expected Wage & 0.07552 & 0.00975 & 0.06463 & 0.00312 & 0.00000 & 0.00001 & 0.32650 & 0.01000 \\

\addlinespace

Location Controls
& \multicolumn{2}{c}{Yes}
& \multicolumn{2}{c}{Yes}
& \multicolumn{2}{c}{Yes}
& \multicolumn{2}{c}{Yes} \\

Race Controls
& \multicolumn{2}{c}{Yes}
& \multicolumn{2}{c}{Yes}
& \multicolumn{2}{c}{Yes}
& \multicolumn{2}{c}{Yes} \\

Age Controls
& \multicolumn{2}{c}{Yes}
& \multicolumn{2}{c}{Yes}
& \multicolumn{2}{c}{Yes}
& \multicolumn{2}{c}{Yes} \\

\bottomrule
\end{tabular}

\begin{tablenotes}[flushleft]
\scriptsize
\item \textit{Notes:} This table reports average marginal effects for the major choice probabilities obtained from the structural estimation in Table~\ref{tab: MNP_me}. Standard errors are computed using a nonparametric bootstrap with 1,000 replications.
\end{tablenotes}

\end{threeparttable}
\end{table}

\begin{table}[htbp]
\centering
\caption{Marginal Effects from Structural Estimation: Major Choice}
\label{tab: MNP_me_marginalchoice_gender}

\scriptsize
\setlength{\tabcolsep}{5pt}

\begin{threeparttable}

\begin{tabular}{L{3cm}C{1.2cm}C{1.2cm}C{1.2cm}C{1.2cm}C{1.2cm}C{1.2cm}C{1.2cm}C{1.2cm}}
\toprule
& \multicolumn{2}{c}{Education}
& \multicolumn{2}{c}{Business}
& \multicolumn{2}{c}{Health}
& \multicolumn{2}{c}{Comp.\ \& Eng.} \\
\cmidrule(lr){2-3}\cmidrule(lr){4-5}\cmidrule(lr){6-7}\cmidrule(lr){8-9}
& AME & Std.\ Err.
& AME & Std.\ Err.
& AME & Std.\ Err.
& AME & Std.\ Err. \\
\midrule

\multicolumn{9}{l}{\textit{Panel A: Female}} \\

\textbf{Marriage Exp.} & 0.00187 & 0.00034 & 0.00081 & 0.00010 & 0.00000 & 0.00000 & -0.01652 & 0.00052 \\
GPA Term 1  & 0.01561 & 0.00298 & -0.00172 & 0.00037 & -0.00001 & 0.00000 & -0.01890 & 0.00264 \\
GPA Term 2  & 0.01026 & 0.00195 & -0.01662 & 0.00092 & -0.00001 & 0.00001 & 0.01094 & 0.00247 \\
GPA Term 3  & 0.01573 & 0.00301 & -0.00863 & 0.00053 & 0.00000 & 0.00000 & -0.01503 & 0.00275 \\
Expected Wage & 0.06240 & 0.01195 & 0.16260 & 0.00936 & 0.00007 & 0.00004 & 0.47118 & 0.01074 \\

\addlinespace
\multicolumn{9}{l}{\textit{Panel B: Male}} \\

\textbf{Marriage Exp.} & 0.00000 & 0.00000 & -0.00014 & 0.00001 & 0.00000 & 0.00000 & -0.00052 & 0.00003 \\
GPA Term 1  & 0.00000 & 0.00000 & -0.00176 & 0.00014 & 0.00000 & 0.00000 & 0.00699 & 0.00033 \\
GPA Term 2  & 0.00000 & 0.00000 & -0.00202 & 0.00015 & 0.00000 & 0.00000 & 0.01712 & 0.00074 \\
GPA Term 3  & 0.00000 & 0.00000 & 0.00288 & 0.00023 & 0.00000 & 0.00000 & -0.02372 & 0.00104 \\
Expected Wage & 0.00001 & 0.00001 & 0.01548 & 0.00119 & 0.00000 & 0.00000 & 0.11087 & 0.00486 \\

\midrule

Location Controls
& \multicolumn{2}{c}{Yes}
& \multicolumn{2}{c}{Yes}
& \multicolumn{2}{c}{Yes}
& \multicolumn{2}{c}{Yes} \\

Race Controls
& \multicolumn{2}{c}{Yes}
& \multicolumn{2}{c}{Yes}
& \multicolumn{2}{c}{Yes}
& \multicolumn{2}{c}{Yes} \\

Age Controls
& \multicolumn{2}{c}{Yes}
& \multicolumn{2}{c}{Yes}
& \multicolumn{2}{c}{Yes}
& \multicolumn{2}{c}{Yes} \\

\bottomrule
\end{tabular}

\begin{tablenotes}[flushleft]
\scriptsize
\item \textit{Notes:} This table reports average marginal effects for major choice probabilities obtained from the structural estimation in Table~\ref{tab: MNPstructuralgender}. Standard errors are computed using a nonparametric bootstrap with 1,000 replications.
\end{tablenotes}

\end{threeparttable}
\end{table}

\newpage

\subsection{Validation Exercise}
\label{sec:appendixvalidation}

Table \ref{tab: MNL_reduced1} presents a descriptive analysis with Multinomial Logistic estimation (MNL) result that includes the stated marriage expectations measure in addition to other previously used controls. The findings indicate that the stated marriage expectations measure has a statistically significant impact on students' college major choices. More specifically, students who have higher self-assessed probabilities of getting married in the next five years are more inclined to choose Education programs and less likely to choose Business programs.

\begin{table}[htbp]
\centering
\caption{Descriptive Regression with Stated Marriage Expectation Measure}
\label{tab: MNL_reduced1}
\scriptsize
\setlength{\tabcolsep}{5pt}

\begin{threeparttable}

\begin{tabular}{L{3cm}C{1.2cm}C{1.2cm}C{1.2cm}C{1.2cm}C{1.2cm}C{1.2cm}C{1.2cm}C{1.2cm}}
\toprule
& \multicolumn{2}{c}{Education}
& \multicolumn{2}{c}{Business}
& \multicolumn{2}{c}{Health}
& \multicolumn{2}{c}{Comp.\ \& Eng.} \\
\cmidrule(lr){2-3}\cmidrule(lr){4-5}\cmidrule(lr){6-7}\cmidrule(lr){8-9}

& Estimate & Std.\ Err.
& Estimate & Std.\ Err.
& Estimate & Std.\ Err.
& Estimate & Std.\ Err. \\
\midrule

\textbf{Marriage Exp.} & 0.008 & 0.004 & -0.005 & 0.003 & -0.002 & 0.004 & 0.002 & 0.004 \\

Female & 1.157 & 0.284 & -0.281 & 0.168 & 0.899 & 0.290 & -2.057 & 0.304 \\

GPA Term 1 & -0.129 & 0.119 & -0.034 & 0.088 & 0.105 & 0.126 & 0.111 & 0.139 \\

GPA Term 2 & 0.188 & 0.144 & -0.036 & 0.095 & -0.137 & 0.129 & 0.363 & 0.167 \\

GPA Term 3 & 0.227 & 0.139 & 0.037 & 0.087 & 0.131 & 0.123 & -0.184 & 0.130 \\

\addlinespace

Expected Earnings
& \multicolumn{2}{c}{Yes}
& \multicolumn{2}{c}{Yes}
& \multicolumn{2}{c}{Yes}
& \multicolumn{2}{c}{Yes} \\

Location Controls
& \multicolumn{2}{c}{Yes}
& \multicolumn{2}{c}{Yes}
& \multicolumn{2}{c}{Yes}
& \multicolumn{2}{c}{Yes} \\

Race Controls
& \multicolumn{2}{c}{Yes}
& \multicolumn{2}{c}{Yes}
& \multicolumn{2}{c}{Yes}
& \multicolumn{2}{c}{Yes} \\

Age Controls
& \multicolumn{2}{c}{Yes}
& \multicolumn{2}{c}{Yes}
& \multicolumn{2}{c}{Yes}
& \multicolumn{2}{c}{Yes} \\

\bottomrule
\end{tabular}

\begin{tablenotes}[flushleft]
\scriptsize
\item \textit{Notes:} This table reports multinomial logit regression estimates of students' major choices including the stated marriage expectation measure. The coefficient on expected earnings is normalized to one.
\end{tablenotes}

\end{threeparttable}
\end{table}

Table \ref{tab: MNP_me} presents estimation results obtained through our proposed estimation method. The estimated coefficients show that stated marriage expectation measure have significant impact on Education and Business major choices. Marginal effect estimates obtained from these estimation results are available in Table \ref{tab: MNP_me_marginalchoice}.

\begin{table}[htbp]
\centering
\caption{Structural Estimation Results: Major Choice}
\label{tab: MNP_me}

\scriptsize
\setlength{\tabcolsep}{5pt}

\begin{threeparttable}

\begin{tabular}{L{3cm}C{1.2cm}C{1.2cm}C{1.2cm}C{1.2cm}C{1.2cm}C{1.2cm}C{1.2cm}C{1.2cm}}
\toprule
& \multicolumn{2}{c}{Education}
& \multicolumn{2}{c}{Business}
& \multicolumn{2}{c}{Health}
& \multicolumn{2}{c}{Comp.\ \& Eng.} \\
\cmidrule(lr){2-3}\cmidrule(lr){4-5}\cmidrule(lr){6-7}\cmidrule(lr){8-9}

& Estimate & Std.\ Err.
& Estimate & Std.\ Err.
& Estimate & Std.\ Err.
& Estimate & Std.\ Err. \\
\midrule

\textbf{Marriage Exp.} & 0.005 & 0.002 & -0.003 & 0.001 & 0.000 & 0.001 & 0.000 & 0.002 \\

Female & 0.717 & 0.099 & -0.240 & 0.079 & 0.291 & 0.098 & -1.258 & 0.099 \\

GPA Term 1 & -0.104 & 0.049 & -0.008 & 0.042 & 0.049 & 0.047 & 0.065 & 0.050 \\

GPA Term 2 & 0.114 & 0.057 & -0.013 & 0.045 & -0.044 & 0.051 & 0.187 & 0.059 \\

GPA Term 3 & 0.131 & 0.053 & 0.024 & 0.046 & 0.071 & 0.049 & -0.081 & 0.053 \\

\addlinespace

Expected Earnings
& \multicolumn{2}{c}{Yes}
& \multicolumn{2}{c}{Yes}
& \multicolumn{2}{c}{Yes}
& \multicolumn{2}{c}{Yes} \\

Location Controls
& \multicolumn{2}{c}{Yes}
& \multicolumn{2}{c}{Yes}
& \multicolumn{2}{c}{Yes}
& \multicolumn{2}{c}{Yes} \\

Race Controls
& \multicolumn{2}{c}{Yes}
& \multicolumn{2}{c}{Yes}
& \multicolumn{2}{c}{Yes}
& \multicolumn{2}{c}{Yes} \\

Age Controls
& \multicolumn{2}{c}{Yes}
& \multicolumn{2}{c}{Yes}
& \multicolumn{2}{c}{Yes}
& \multicolumn{2}{c}{Yes} \\

\bottomrule
\end{tabular}

\begin{tablenotes}[flushleft]
\scriptsize
\item \textit{Notes:} This table reports the college major choice component of the structural estimation for the full sample. Estimation results for the marriage outcome equation are presented in Table~\ref{tab: corr_eq}.
\end{tablenotes}

\end{threeparttable}
\end{table}

\subsection{Heterogeneous Effects}

\subsubsection{Descriptive Analysis}

Tables \ref{tab: MNL_reduced_gender} and \ref{tab:results22} present descriptive evidence for the disparities between male and female students major choice and marriage outcomes. From these regressions, we can infer that higher early college GPA
leads to different major choice between male and female students. Moreover, a gender-based marital status disparity is evident in the descriptive regressions: female graduates show a higher likelihood of getting married in comparison to their male counterparts, even after controlling for college-major choice and other relevant factors. The descriptive analysis also underscores the noteworthy impact of average earnings on observed marital status, revealing a significant and positive relationship between the two. However, the effect of
average working hours on marital status is found to be statistically insignificant.

\begin{table}[htbp]
\centering
\caption{Descriptive Regressions by Gender: Major Choice}
\label{tab: MNL_reduced_gender}

\scriptsize
\setlength{\tabcolsep}{5pt}

\begin{threeparttable}

\begin{tabular}{L{3cm}C{1.2cm}C{1.2cm}C{1.2cm}C{1.2cm}C{1.2cm}C{1.2cm}C{1.2cm}C{1.2cm}}
\toprule
& \multicolumn{2}{c}{Education}
& \multicolumn{2}{c}{Business}
& \multicolumn{2}{c}{Health}
& \multicolumn{2}{c}{Comp.\ \& Eng.} \\
\cmidrule(lr){2-3}\cmidrule(lr){4-5}\cmidrule(lr){6-7}\cmidrule(lr){8-9}

& Estimate & Std.\ Err.
& Estimate & Std.\ Err.
& Estimate & Std.\ Err.
& Estimate & Std.\ Err. \\
\midrule

\multicolumn{9}{l}{\textit{Panel A: Female }} \\

GPA Term 1 & 0.037 & 0.147 & 0.075 & 0.116 & 0.158 & 0.136 & 0.161 & 0.285 \\
GPA Term 2 & 0.099 & 0.159 & -0.073 & 0.119 & -0.164 & 0.135 & 0.077 & 0.307 \\
GPA Term 3 & 0.267 & 0.158 & -0.009 & 0.110 & 0.112 & 0.129 & 0.073 & 0.280 \\

\addlinespace
\multicolumn{9}{l}{\textit{Panel B: Male }} \\

GPA Term 1 & -0.551 & 0.213 & -0.175 & 0.126 & -0.028 & 0.298 & 0.063 & 0.156 \\
GPA Term 2 & 0.328 & 0.320 & 0.047 & 0.146 & 0.141 & 0.332 & 0.419 & 0.188 \\
GPA Term 3 & 0.268 & 0.285 & 0.098 & 0.133 & -0.055 & 0.286 & -0.210 & 0.146 \\

\midrule

Expected Earnings
& \multicolumn{2}{c}{Yes}
& \multicolumn{2}{c}{Yes}
& \multicolumn{2}{c}{Yes}
& \multicolumn{2}{c}{Yes} \\

Location Controls
& \multicolumn{2}{c}{Yes}
& \multicolumn{2}{c}{Yes}
& \multicolumn{2}{c}{Yes}
& \multicolumn{2}{c}{Yes} \\

Race Controls
& \multicolumn{2}{c}{Yes}
& \multicolumn{2}{c}{Yes}
& \multicolumn{2}{c}{Yes}
& \multicolumn{2}{c}{Yes} \\

Age Controls
& \multicolumn{2}{c}{Yes}
& \multicolumn{2}{c}{Yes}
& \multicolumn{2}{c}{Yes}
& \multicolumn{2}{c}{Yes} \\

\bottomrule
\end{tabular}

\begin{tablenotes}[flushleft]
\scriptsize
\item \textit{Notes:} This table reports multinomial logit regression estimates of students' major choices estimated separately for female and male students. The coefficient on normalized expected earnings is fixed at one. Remaining majors are grouped as ``Other'' and serve as the base category.
\end{tablenotes}

\end{threeparttable}
\end{table}

\begin{table}[htbp]
\centering
\caption{Descriptive Regressions by Gender: Marriage Outcome}
\label{tab:results22}

\scriptsize
\setlength{\tabcolsep}{5pt}

\begin{threeparttable}

\begin{tabular}{L{3.2cm}C{1.2cm}C{1.2cm}C{1.2cm}C{1.2cm}C{1.2cm}C{1.2cm}}
\toprule
& \multicolumn{2}{c}{(1)} & \multicolumn{2}{c}{(2)} & \multicolumn{2}{c}{(3)} \\
\cmidrule(lr){2-3}\cmidrule(lr){4-5}\cmidrule(lr){6-7}
& Estimate & Std.\ Err. & Estimate & Std.\ Err. & Estimate & Std.\ Err. \\
\midrule

\multicolumn{7}{l}{\textit{Panel A: Female}} \\

Education & 1.385 & 0.231 & 1.257 & 0.253 & 1.571 & 0.287 \\
Business & 1.156 & 0.190 & 1.010 & 0.224 & 1.277 & 0.252 \\
Health & 1.031 & 0.205 & 0.884 & 0.238 & 1.139 & 0.262 \\
Comp.\ \& Eng. & 1.111 & 0.352 & 0.924 & 0.384 & 1.185 & 0.399 \\
Other & 0.992 & 0.154 & 0.865 & 0.185 & 1.127 & 0.217 \\
Earnings &  &  & 0.004 & 0.003 & 0.009 & 0.004 \\
Working Hours &  &  &  &  & -0.236 & 0.100 \\

\addlinespace
\multicolumn{7}{l}{\textit{Panel B: Male}} \\

Education & 0.985 & 0.334 & 0.621 & 0.354 & 0.547 & 0.369 \\
Business & 0.714 & 0.194 & 0.263 & 0.242 & 0.175 & 0.270 \\
Health & 1.164 & 0.358 & 0.717 & 0.388 & 0.637 & 0.402 \\
Comp.\ \& Eng. & 0.878 & 0.206 & 0.376 & 0.260 & 0.301 & 0.280 \\
Other & 0.877 & 0.157 & 0.458 & 0.206 & 0.375 & 0.235 \\
Earnings &  &  & 0.011 & 0.004 & 0.010 & 0.004 \\
Working Hours &  &  &  &  & 0.062 & 0.084 \\

\midrule

Location Controls & \multicolumn{2}{c}{Yes} & \multicolumn{2}{c}{Yes} & \multicolumn{2}{c}{Yes} \\
Race Controls & \multicolumn{2}{c}{Yes} & \multicolumn{2}{c}{Yes} & \multicolumn{2}{c}{Yes} \\
Age Controls & \multicolumn{2}{c}{Yes} & \multicolumn{2}{c}{Yes} & \multicolumn{2}{c}{Yes} \\

\bottomrule
\end{tabular}

\begin{tablenotes}[flushleft]
\scriptsize
\item \textit{Notes:} This table reports probit regression estimates for marital status separately for female and male college graduates under three alternative covariate specifications. Panel A presents estimates for female graduates, and Panel B presents estimates for male graduates.
\end{tablenotes}

\end{threeparttable}
\end{table}

\subsubsection{Structural Analysis}

We apply the proposed estimation method separately to females and males as outlined in Section \ref{testing}, omitting the stated marriage expectation measure. The primary parameters of interest are the significance of the vector correlation coefficients $\boldsymbol{\rho}$. As shown in Table \ref{tab:FM_nome}, some elements of the $\boldsymbol{\rho}$ vectors are statistically different from zero for both female and male graduates ($\rho_5$ for females; $\rho_4$ and $\rho_5$ for males), suggesting that the unobserved factors in the equations for marital outcomes and college major choices are correlated when stated marriage expectations are not controlled for.\footnote{The estimation results related to the major choice component are omitted for brevity.}

\begin{table}[htbp]
\centering
\caption{Structural Estimation Results by Gender: Marriage Outcome}
\label{tab:FM_nome}

\scriptsize
\setlength{\tabcolsep}{6pt}

\begin{threeparttable}

\begin{tabular}{L{4.2cm}C{1.6cm}C{1.6cm}C{1.6cm}C{1.6cm}}
\toprule
& \multicolumn{2}{c}{Female} & \multicolumn{2}{c}{Male} \\
\cmidrule(lr){2-3}\cmidrule(lr){4-5}
& Estimate & Std.\ Err. & Estimate & Std.\ Err. \\
\midrule

\multicolumn{5}{l}{\textit{Panel A: Marriage Equation}} \\

Education & 1.817 & 0.455 & 1.999 & 0.738 \\
Business & 1.542 & 0.610 & 1.068 & 0.287 \\
Health & 1.385 & 0.417 & 3.908 & 0.302 \\
Comp.\ \& Eng. & 1.675 & 1.070 & 1.630 & 0.333 \\
Other & 1.408 & 0.300 & -0.325 & 0.247 \\
Avg.\ Observed Earnings & 0.049 & 0.097 & 0.008 & 0.060 \\
Avg.\ Observed Working Hours & -0.186 & 0.110 & 0.201 & 0.058 \\

\addlinespace
\multicolumn{5}{l}{\textit{Panel B: Dependence Parameters}} \\

$\rho_{1}$ & 0.027 & 0.140 & -0.498 & 0.353 \\
$\rho_{2}$ & 0.105 & 0.238 & -0.399 & 0.426 \\
$\rho_{3}$ & 0.062 & 0.141 & -1.000 & 0.998 \\
$\rho_{4}$ & 0.058 & 0.273 & -0.815 & 0.291 \\
$\rho_{5}$ & 0.029 & 0.001 & 0.897 & 0.316 \\

\midrule

Location Controls & \multicolumn{2}{c}{Yes} & \multicolumn{2}{c}{Yes} \\
Race Controls & \multicolumn{2}{c}{Yes} & \multicolumn{2}{c}{Yes} \\
Age Controls & \multicolumn{2}{c}{Yes} & \multicolumn{2}{c}{Yes} \\

\bottomrule
\end{tabular}

\begin{tablenotes}[flushleft]
\scriptsize
\item \textit{Notes:} This table reports structural estimation results for the marriage outcome equation and the dependence parameters estimated separately for female and male graduates. Choice probabilities are computed using 250 GHK simulations. Standard errors are calculated from the outer product of the log-likelihood gradient. The correlation parameters $\boldsymbol{\rho}$ are estimated using the transformation
$2(2/(1+exp(\rho)) -1)$.
\end{tablenotes}

\end{threeparttable}
\end{table}

\begin{table}[htbp]
\centering
\caption{Multinomial Logit Results by Gender: Major Choice}
\label{tab: MNL_reduced1_gender}

\scriptsize
\setlength{\tabcolsep}{5pt}

\begin{threeparttable}

\begin{tabular}{L{3cm}C{1.2cm}C{1.2cm}C{1.2cm}C{1.2cm}C{1.2cm}C{1.2cm}C{1.2cm}C{1.2cm}}
\toprule
& \multicolumn{2}{c}{Education}
& \multicolumn{2}{c}{Business}
& \multicolumn{2}{c}{Health}
& \multicolumn{2}{c}{Comp.\ \& Eng.} \\
\cmidrule(lr){2-3}\cmidrule(lr){4-5}\cmidrule(lr){6-7}\cmidrule(lr){8-9}

& Estimate & Std.\ Err.
& Estimate & Std.\ Err.
& Estimate & Std.\ Err.
& Estimate & Std.\ Err. \\
\midrule

\multicolumn{9}{l}{\textit{Panel A: Female}} \\

\textbf{Marriage Exp.} & 0.008 & 0.005 & -0.007 & 0.004 & -0.004 & 0.004 & -0.010 & 0.009 \\
GPA Term 1 & 0.014 & 0.147 & 0.078 & 0.119 & 0.162 & 0.141 & 0.117 & 0.286 \\
GPA Term 2 & 0.148 & 0.169 & -0.090 & 0.127 & -0.208 & 0.143 & 0.068 & 0.325 \\
GPA Term 3 & 0.227 & 0.158 & -0.001 & 0.116 & 0.162 & 0.137 & 0.069 & 0.293 \\

\addlinespace
\multicolumn{9}{l}{\textit{Panel B: Male}} \\

\textbf{Marriage Exp.} & 0.007 & 0.009 & -0.003 & 0.004 & 0.008 & 0.009 & 0.007 & 0.005 \\
GPA Term 1 & -0.511 & 0.217 & -0.187 & 0.131 & -0.023 & 0.297 & 0.093 & 0.159 \\
GPA Term 2 & 0.311 & 0.331 & 0.046 & 0.150 & 0.196 & 0.354 & 0.425 & 0.194 \\
GPA Term 3 & 0.225 & 0.292 & 0.096 & 0.137 & -0.074 & 0.299 & -0.234 & 0.150 \\

\midrule

Expected Earnings
& \multicolumn{2}{c}{Yes}
& \multicolumn{2}{c}{Yes}
& \multicolumn{2}{c}{Yes}
& \multicolumn{2}{c}{Yes} \\

Location Controls
& \multicolumn{2}{c}{Yes}
& \multicolumn{2}{c}{Yes}
& \multicolumn{2}{c}{Yes}
& \multicolumn{2}{c}{Yes} \\

Race Controls
& \multicolumn{2}{c}{Yes}
& \multicolumn{2}{c}{Yes}
& \multicolumn{2}{c}{Yes}
& \multicolumn{2}{c}{Yes} \\

Age Controls
& \multicolumn{2}{c}{Yes}
& \multicolumn{2}{c}{Yes}
& \multicolumn{2}{c}{Yes}
& \multicolumn{2}{c}{Yes} \\

\bottomrule
\end{tabular}

\begin{tablenotes}[flushleft]
\scriptsize
\item \textit{Notes:} This table reports multinomial logit regression estimates of students' major choices including the stated marriage expectation measure. Panel A reports estimates for female students and Panel B reports estimates for male students. The coefficient on expected earnings is normalized to one.
\end{tablenotes}

\end{threeparttable}
\end{table}

Table \ref{tab: MNL_reduced1_gender} presents descriptive regressions of major choices for female and male students separately. These regressions incorporate the stated marriage expectation measure, serving both as preliminary evidence and as a benchmark for our structural estimation results. The descriptive regressions offer model-free insights into the effect of marriage preferences on major choices. The structural regression results for major choices, presented in Table \ref{tab: MNPstructuralgender}, reveal similar patterns. Students with a higher self-assessed probability of marrying within the next five years are more likely to choose Education programs and less likely to opt for Business, Health, or Computer \& Engineering programs. However, the magnitude of these effects varies depending on the estimation method used.

\begin{table}[htbp]
\centering
\caption{Structural Estimation Results by Gender: Major Choice}
\label{tab: MNPstructuralgender}

\scriptsize
\setlength{\tabcolsep}{5pt}

\begin{threeparttable}

\begin{tabular}{L{3cm}C{1.2cm}C{1.2cm}C{1.2cm}C{1.2cm}C{1.2cm}C{1.2cm}C{1.2cm}C{1.2cm}}
\toprule
& \multicolumn{2}{c}{Education}
& \multicolumn{2}{c}{Business}
& \multicolumn{2}{c}{Health}
& \multicolumn{2}{c}{Comp.\ \& Eng.} \\
\cmidrule(lr){2-3}\cmidrule(lr){4-5}\cmidrule(lr){6-7}\cmidrule(lr){8-9}

& Estimate & Std.\ Err.
& Estimate & Std.\ Err.
& Estimate & Std.\ Err.
& Estimate & Std.\ Err. \\
\midrule

\multicolumn{9}{l}{\textit{Panel A: Female (N = 654)}} \\

\textbf{Marriage Exp.} & 0.014 & 0.007 & -0.011 & 0.005 & -0.039 & 0.007 & -0.037 & 0.007 \\
GPA Term 1 & 0.252 & 0.101 & 0.018 & 0.084 & -0.105 & 0.112 & -0.005 & 0.185 \\
GPA Term 2 & 0.149 & 0.102 & -0.082 & 0.097 & -0.193 & 0.109 & 0.013 & 0.198 \\
GPA Term 3 & 0.242 & 0.118 & -0.026 & 0.111 & -0.019 & 0.111 & -0.011 & 0.186 \\

\addlinespace
\multicolumn{9}{l}{\textit{Panel B: Male (N = 527)}} \\

\textbf{Marriage Exp.} & 0.029 & 0.003 & -0.014 & 0.002 & -0.064 & 0.003 & -0.008 & 0.002 \\
GPA Term 1 & -0.191 & 0.080 & -0.095 & 0.062 & -0.054 & 0.091 & 0.038 & 0.060 \\
GPA Term 2 & 0.177 & 0.131 & -0.025 & 0.070 & -0.021 & 0.095 & 0.149 & 0.068 \\
GPA Term 3 & 0.136 & 0.116 & 0.058 & 0.062 & -0.108 & 0.080 & -0.198 & 0.060 \\

\midrule

Expected Earnings
& \multicolumn{2}{c}{Yes}
& \multicolumn{2}{c}{Yes}
& \multicolumn{2}{c}{Yes}
& \multicolumn{2}{c}{Yes} \\

Location Controls
& \multicolumn{2}{c}{Yes}
& \multicolumn{2}{c}{Yes}
& \multicolumn{2}{c}{Yes}
& \multicolumn{2}{c}{Yes} \\

Race Controls
& \multicolumn{2}{c}{Yes}
& \multicolumn{2}{c}{Yes}
& \multicolumn{2}{c}{Yes}
& \multicolumn{2}{c}{Yes} \\

Age Controls
& \multicolumn{2}{c}{Yes}
& \multicolumn{2}{c}{Yes}
& \multicolumn{2}{c}{Yes}
& \multicolumn{2}{c}{Yes} \\

\midrule
Observations & \multicolumn{8}{c}{1181} \\
\bottomrule

\end{tabular}

\begin{tablenotes}[flushleft]
\scriptsize
\item \textit{Notes:} This table presents the college major choice component of the proposed structural estimation separately for female and male graduates. Panel A reports estimates for female graduates and Panel B reports estimates for male graduates. Estimation results for the marriage outcome component of the joint estimation are reported in Table \ref{tab: selection_eq_gender}.
\end{tablenotes}

\end{threeparttable}
\end{table}

\clearpage

\subsection{Additional Figures}

\begin{figure}[ht]
{\centering
\caption{Majors of college graduates in the US, 2001-2013}
\label{fig:pop_majors}
\includegraphics[scale=0.6]{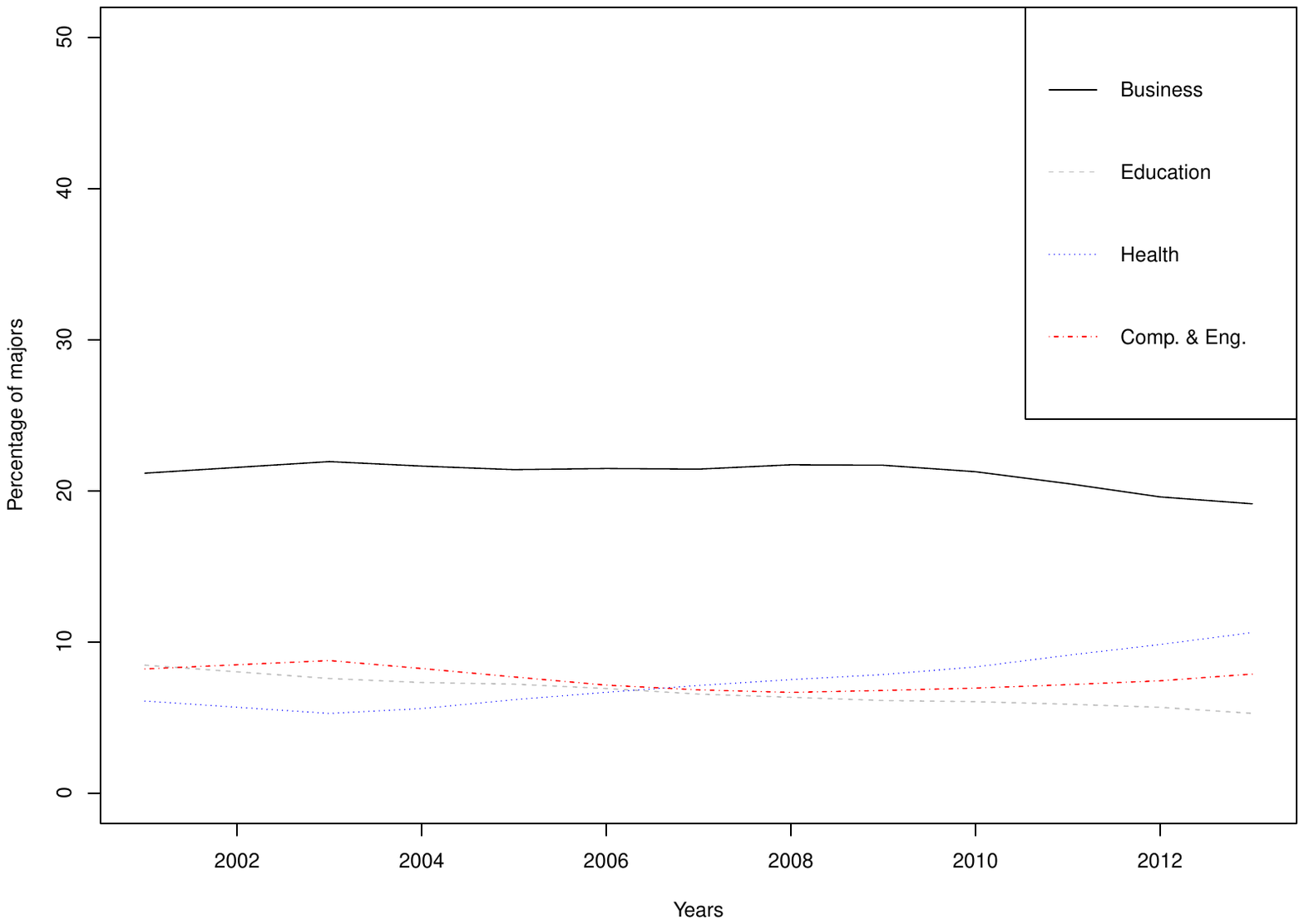}}
\scriptsize{\textit{Notes:} This figure presents percentages of selected college majors of college graduates of the NLSY97 sample for the years between 2001 and 2013 in the US.}
\end{figure}

\end{document}